\documentclass[12pt,a4paper]{article}

\usepackage[margin=2.5cm]{geometry}
\usepackage{graphicx}
\usepackage{amsmath}
\usepackage{nicefrac}
\usepackage{chngcntr}
	\counterwithout{footnote}{section}
\usepackage[bottom]{footmisc}
\usepackage{booktabs}
\usepackage{multirow}
\usepackage{tabularx}
\usepackage[raiselinks=true,
						bookmarks=true,
						bookmarksopenlevel=1,
						bookmarksopen=true,
						bookmarksnumbered=true,
						hyperindex=true,
						hypertexnames=false,
  						plainpages=false,
						pdfpagelabels=true,
						pdfborder={0 0 0.5},
						colorlinks=false,						
						linkbordercolor={0 0.61 0.50},   
						citebordercolor={0 0.61 0.50},
						pdfstartview=FitB]{hyperref}

\title{Corrections of Order $\alpha \, \alpha_s$ to $W$ Boson Decays}
\author{Dominik Kara \\\\
	\textit{Institut f\"ur Theoretische Teilchenphysik} \\
	\textit{Karlsruhe Institute of Technology (KIT)} \\
	\textit{76128 Karlsruhe, Germany}
	}

\date{}

\begin{document}

\hfill TTP13-026
{\let\newpage\relax\maketitle}
\thispagestyle{empty}

\begin{abstract}
We present the calculation of the mixed two-loop QCD/electroweak corrections to hadronic \mbox{$W$ boson} decays within the Standard Model. The optical theorem is applied to the $W$ boson two-point function. The multi-scale integrals are computed with the help of asymptotic expansions, which factorize the three-loop diagrams into one- and two-loop vacuum and propagator-type integrals.
\end{abstract}

\section{Introduction}
\label{sec:Intro}

The discovery of the $W$ boson at the CERN Super Proton Synchrotron ($Sp\bar{p}S$) collider in 1983 \cite{Arnison:1983, Banner:1983} set a milestone for the success of the Standard Model. At the Large Electron-Positron Collider (LEP), pairs of $W$ bosons were produced and the $W$ mass was determined to high accuracy \cite{Alcaraz:2006}. At  Tevatron, measurements improved \cite{Group:2009} and have led to the current world average of the $W$ boson mass and the total decay width as well as to the hadronic branching ratio \cite{Beringer:2012}:
\begin{align}
	M_W &= (80.385 \pm 0.015) \, \mathrm{GeV} \notag \, ,\\
	\Gamma_\mathrm{tot} &= (2.085 \pm 0.042) \, \mathrm{GeV} \notag \, ,\\
	\mathrm{BR} (W \rightarrow \mathrm{hadrons}) &= (67.60 \pm 0.27) \% \, .
\end{align}
The $W$ boson is connected with the top quark and the Higgs boson via radiative corrections. Hence, precise measurements of the top and $W$ masses allowed constraining the mass range of the Higgs boson in the past. In addition, the hadronic partial widths can serve to determine the elements of the quark mixing matrix \cite{Longo:1986}.\\
In order to compare these experimental results to theoretical predictions, we need calculations of increased accuracy. In our case, this requires the computation of the mixed QCD/electroweak corrections to the two-particle decay of the $W^+$ boson into quarks. Thus, we consider $\mathcal{O}(\alpha \, \alpha_s)$ corrections to the decay width of the process
\begin{equation}
W^+ \rightarrow q \, \bar{q'}
\end{equation}
within the Standard Model, where $q = u, c$ and $q' = d, s, b$ stand for any charge preserving combination of up- and down-type quarks in the final state. The result for the charge-conjugate process $W^- \rightarrow \bar{q} \, q'$ will be the same.\\
Up to now, the following corrections to hadronic $W$ boson decays have been calculated on the theoretical side:
\begin{itemize}
\item	One-loop QED corrections for massless fermions \cite{Marciano:1974, Albert:1980}
\item	One-loop electroweak corrections for massless fermions \cite{Inoue:1980, Consoli:1983, Bardin:1986, Jegerlehner:1986}
\item	One-loop QCD corrections for finite quark masses \cite{Chang:1982, Alvarez:1988}
\item	One-loop electroweak and QCD corrections for finite fermion masses \cite{Sack:1990, Kniehl:2000}
\item	Two- and three-loop QCD corrections for massless quarks \cite{Gorishnii:1990, Surguladze:1991, Chetyrkin:1997}
\item	Two- and three-loop QCD corrections including quadratic quark mass corrections~\cite{Kuhn:1997}
\item Four-loop QCD corrections for massless quarks \cite{Baikov:2008}
\end{itemize}
In addition, we would like to mention Ref. \cite{Czarnecki:1996}. Therein, the two-loop mixed QCD/elec\-tro\-weak corrections were calculated for the decay of the $Z$ boson into the light quark flavors $u$, $d$, $s$ and $c$.\footnote{Refs. \cite{Seidensticker:1998, Harlander:1997, Fleischer:1999} extended this calculation by applying the optical theorem and asymptotic expansions in $M_W^2 \ll M_t^2$ to the decay of the $Z$ boson into massive bottom quarks.} We have approached the $W$ boson decay using the same techniques, namely the optical theorem and asymptotic expansions.\\
The optical theorem relates the decay width to the transversal part of the $W$ boson two-point function,
\begin{equation}
	\Gamma = \frac{1}{M_W} \, \mathrm{Im} \, \Pi^W_T(M_W^2) \, ,
\end{equation}
leading to multi-scale three-loop two-point functions. Their exact computation requires great efforts and is not thoroughly understood in contrast to the calculation resulting from the application of asymptotic expansions, which provide a well-studied, systematic expansion in heavy masses or large momenta. The results of this method have turned out to agree with the leading-order terms of the exact results in many cases, i.e. the numerical form of the exact result can be reproduced to high accuracy \cite{Smirnov:2001, Harlander:1998, Harlander:1999, Seidensticker:1999}. By defining a hierarchy of scales, the integrals are split into several parts which are solved separately. In our case, this corresponds to a factorization of three-loop diagrams into one- and two-loop vacuum and propagator-type integrals. Consequently, the result is obtained in powers of $x$ and powers of logarithms of $x$. Here, $x$ is given by $q^2/M^2$ where $q$ denotes the external momentum of the two-point function and $M$ indicates the mass of a boson occurring inside the loop. The general prescription proceeds as follows:
\begin{itemize}
\item	Expand asymptotically in $x \ll 1$. This type of expansion is referred to as the `hard mass procedure'; we will call the associated asymptotic series '$S$-series'.
\item	Carry out the integrations of the various parts.
\item	Obtain the result by approaching $x \rightarrow 1$ provided that the series converges in this limit. This condition is the basis of our calculations.
\end{itemize}
Since $q^2$ and $M^2$ are not considered equal initially, such a calculation is called `off-shell' in contrast to an `on-shell' computation. The expansion can also be performed in $x \gg 1$, which is referred to as the `large momentum procedure'; the associated asymptotic series will be called `$T$-series'.\\

\begin{figure}[tb]
	\begin{center}
	\includegraphics[scale=0.8]{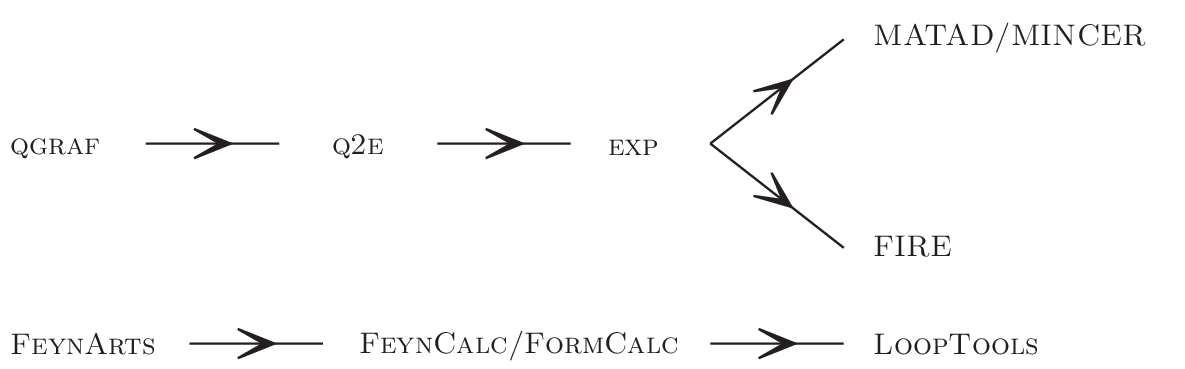}
	\caption[Sequential use of various program packages]{\textbf{Sequential use of various program packages}}
	\label{fig:impl1}
	\end{center}
\end{figure}

In case the series does not converge sufficiently fast, the application of a Pad\'{e} approximation improves the convergence behavior \cite{Fleischer:1994, Broadhurst:1994, Chetyrkin:1996, Broadhurst:1993, Baikov:1995} and allows reproducing the on-shell result via extrapolation. For this purpose, the exact on-shell result has to be calculated using integration-by-parts \cite{Smirnov:2001,Smirnov:2006,Chetyrkin:1981}.\\
These techniques require the consecutive use of several program packages, as shown in Fig. \ref{fig:impl1}. Generally, \textsc{qgraf} \cite{Nogueira:1993} has been applied to generate Feynman diagrams. \textsc{q2e} then transforms the \textsc{qgraf} output into \textsc{exp} readable code \cite{Seidensticker:1999, Seidensticker}. Whenever off-shell calculations are carried out, \textsc{exp} performs asymptotic expansions and maps the expanded expressions on \textsc{MATAD} \cite{Steinhauser:2001} and \textsc{MINCER} \cite{Larin:1991} topologies. These \textsc{Form} \cite{Vermaseren:1992, Vermaseren:2000} programs reduce the remaining integrals to master integrals, which are inserted immediately. In case of an exact on-shell calculation, \textsc{exp} still maps on topologies and \textsc{FIRE} \cite{Smirnov:2008} can be used to reduce the expressions to a set of master integrals. Apart from that, \textsc{FeynArts} \cite{Hahn:2001, Hahn:2010}, \textsc{FormCalc} or \textsc{FeynCalc} \cite{Mertig:1990} and \textsc{LoopTools} \cite{Hahn:2010} have been used to calculate the renormalization constants and to check parts of the results.\\
Throughout all our calculations, we suppose massless quarks and dimensional regularization with $D = 4-2 \epsilon$ dimensions. In addition, we work in Feynman-'t Hooft gauge except for calculations involving gluons, i.e. $\xi_W = \xi_Z = \xi_{\gamma} = 1$. The Cabibbo-Kobayashi-Maskawa (CKM) matrix is taken to be equal to the unit matrix when we calculate higher-order corrections and renormalization constants. However, the different partial decay widths are accounted for by multiplying the Born decay width by the corresponding CKM matrix element squared. It should be mentioned that the contributions to the corrections of $\mathcal{O}(\alpha \, \alpha_s)$ which include non-diagonal CKM matrix elements are largely suppressed. For the electroweak corrections of $\mathcal{O}(\alpha)$, this procedure leads to results which agree with the existing ones for general CKM matrix to high accuracy as detailed in Section~\ref{sec:num}.

\section{Analytical Results}

The hadronic decay width can be decomposed as follows:
\begin{equation}
	\Gamma_\mathrm{had} = \underbrace{\vphantom{\sum_{i=1}^4}\Gamma^{(0)}}_{\text{LO}} + \underbrace{\sum_{i=1}^4 \Gamma^{(i)}_\mathrm{QCD}}_{\text{NLO + HO}} + \underbrace{\vphantom{\sum_{i=1}^4}\Gamma^{(1)}_\mathrm{EW}}_{\text{NLO}} + \underbrace{\vphantom{\sum_{i=1}^4}\Gamma^{(2)}_\mathrm{mixed}}_{\text{NNLO}} \, .
\label{OverallWidth}
\end{equation}
The first two terms involve the leading-order (LO) Born decay width and the next-to-leading-order (NLO) QCD corrections of $\mathcal{O}(\alpha_s)$. The third part describes the electroweak corrections of $\mathcal{O}(\alpha)$ which contribute to the NLO decay width as well. The last term in Eq. \eqref{OverallWidth} stands for the next-to-next-to-leading-order (NNLO) decay width with the so far unknown mixed QCD/electroweak corrections of $\mathcal{O}(\alpha \, \alpha_s)$. The various analytical results will be presented in Sections \ref{sec:ana1}, \ref{sec:ana2} and \ref{sec:ana3}, respectively. The numerical evaluation in Section~\ref{sec:num} will include the higher-order (HO) QCD corrections of $\mathcal{O}(\alpha_s^2)$, $\mathcal{O}(\alpha_s^3)$ and $\mathcal{O}(\alpha_s^4)$ as well. All main formulas are also available as a \textsc{Mathematica} file.\footnote{\url{http://www-ttp.particle.uni-karlsruhe.de/Progdata/ttp13/ttp13-026/}}

\subsection{Leading-Order Born Decay Width and\texorpdfstring{\\}{} Next-To-Leading-Order QCD Corrections}
\label{sec:ana1}

\begin{figure}[b!]
	\begin{center}
	\includegraphics[scale=0.5]{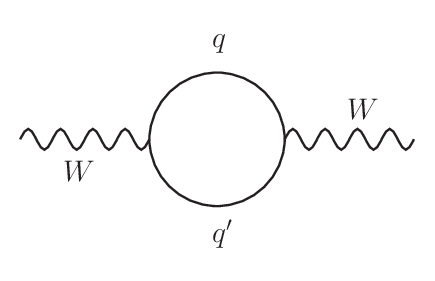}
	\caption[One-loop diagram for the calculation of the Born decay width]{\textbf{One-loop diagram for the calculation of the Born decay width} $\Gamma^{(0)}$}
	\label{fig:born}
	\end{center}
\end{figure}
\begin{figure}[b!]
	\begin{center}
	\includegraphics[scale=0.5]{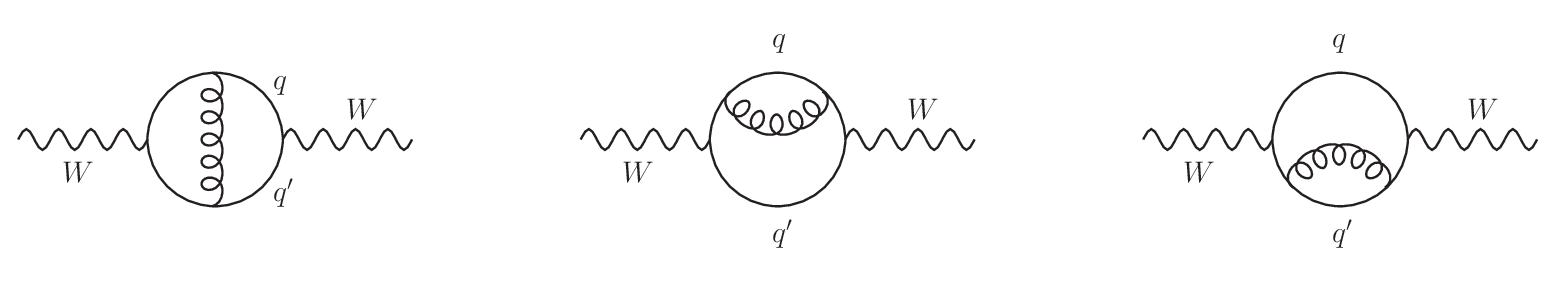}
	\caption[Two-loop diagrams for the calculation of the $\mathcal{O}(\alpha_s)$ QCD corrections]{\textbf{Two-loop diagrams for the calculation of the $\boldsymbol{\mathcal{O}(\alpha_s)}$ QCD corrections} $\Gamma^{(1)}_{\mathrm{QCD}}$. The curly lines stand for gluons.}
	\label{fig:qcd}
	\end{center}
\end{figure}
The Born decay width for the hadronic $W$~boson decay can be reproduced by computing the diagram in Fig. \ref{fig:born}. Expanding the result in $\epsilon$, evaluating the two-point function for $q^2 = M_W^2$ and extracting its imaginary part yields
\begin{equation}
\label{born}
	\Gamma^{(0)} = \frac{\alpha \, M_W \, n_c}{12 \, s_w^2} \, \left| V_{qq'} \right|^2
\end{equation}
with the fine-structure constant $\alpha$, the $W$ boson mass $M_W$, the number of colors in QCD $n_c$, the sine squared of the weak mixing angle $s_w^2 = 1 - c_w^2$ and the CKM matrix element~$V_{qq'}$.\\
For renormalization at NLO, we will make use of the Born decay width up to $\mathcal{O}(\epsilon)$,
\begin{equation}
\Gamma^{(0)}_\epsilon = \Gamma^{(0)} \, \left[1 + \epsilon \left(\frac{5}{3} + \mathrm{ln} \, \frac{\mu^2}{M_W^2} - \mathrm{ln} \, x \right) \right] \, ,
\label{borneps}
\end{equation}
where $x = q^2/M_W^2$.
\clearpage
The calculation of the QCD corrections with the help of Fig. \ref{fig:qcd} proceeds in the same way and results in the finite expression
\begin{equation}
	\Gamma^{(1)}_{\mathrm{QCD}} = \frac{\alpha \, M_W \, (n_c^2 - 1)}{32 \, s_w^2} \cdot \frac{\alpha_s}{\pi} \cdot \left| V_{qq'}\right|^2 \, .
\end{equation}
It yields the well-known QCD correction factor $\alpha_s/\pi$ to the Born decay width for $n_c = 3$:
\begin{equation}
	\Gamma^{(1)}_{\mathrm{QCD}} = \Gamma^{(0)} \cdot \frac{\alpha_s}{\pi} \, .
	\label{qcd}
\end{equation}
As in the LO case, we will need the QCD corrections to the decay width for $n_c = 3$ up to $\mathcal{O}(\epsilon)$ in order to renormalize the results at NNLO:
\begin{equation}
	\Gamma^{(1)}_{\mathrm{QCD},\epsilon} = \Gamma^{(0)} \cdot \frac{\alpha_s}{\pi}\, \left[1 + \epsilon \left(\frac{55}{6} - 8 \, \zeta(3) + 2 \, \mathrm{ln} \, \frac{\mu^2}{M_W^2} - 2 \, \mathrm{ln} \, x \right) \right] \, .
\label{qcdeps}
\end{equation}
$\zeta(s)$ indicates the Riemann zeta function.

\subsection{Next-To-Leading-Order Decay Width:\texorpdfstring{\\}{} Electroweak Corrections of Order \texorpdfstring{$\boldsymbol{\alpha}$}{alpha}}
\label{sec:ana2}

Unlike the large momentum procedure ($q^2 \gg M_W^2$), the hard mass procedure ($q^2 \ll M_W^2$) allows only cuts of massless lines within a Feynman diagram. Hence, the hard mass procedure appropriately describes the properties of the decay $W \rightarrow q \bar{q}'$ since no massive bosons occur in the final state. Following this reasoning, we suppose
\begin{equation}
	x = \frac{q^2}{M_W^2} \ll 1 \, ,
	\label{ewhardmass}
\end{equation}
leading to a result of the form
\begin{equation}
	\label{ewasympnotation}
	S^{(1)} = \Gamma^{(0)} \, \frac{\alpha}{\pi} \, \sum_{n=0}^\infty c_n \, x^n \equiv \Gamma^{(0)} \, \frac{\alpha}{\pi} \, s^{(1)} \, .
\end{equation}
\begin{figure}[tb]
	\begin{center}
	\includegraphics[scale=0.5]{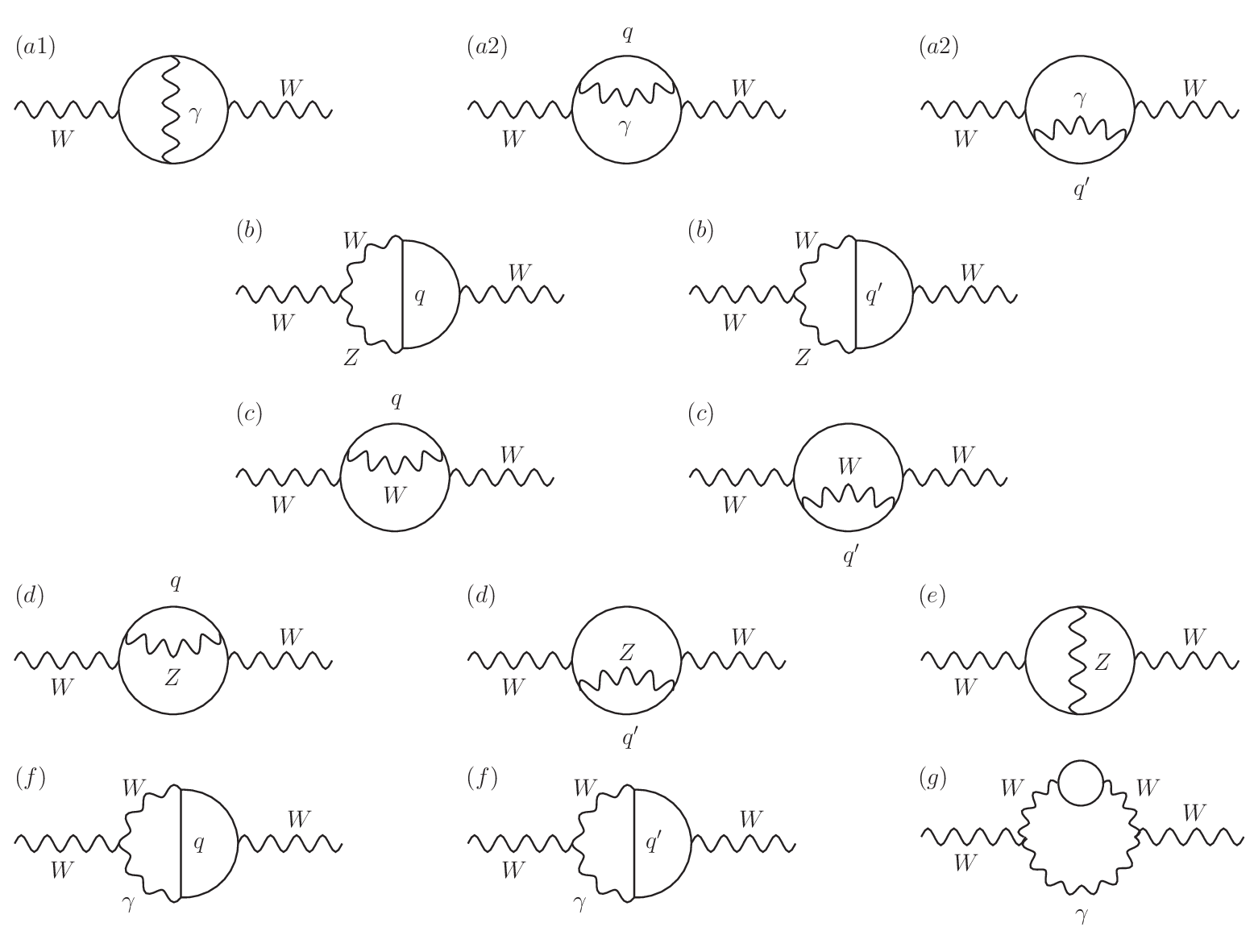}
	\caption[Two-loop diagrams for the calculation of the $\mathcal{O}(\alpha)$ electroweak corrections]{\textbf{Two-loop diagrams for the calculation of the $\boldsymbol{\mathcal{O}(\alpha)}$ electroweak corrections} $\Gamma^{(1)}_{\mathrm{EW}}$. The thirteen diagrams are classified into seven groups $(a)-(g)$. The unlabeled lines stand for quarks.}
	\label{fig:ewdiag}
	\end{center}
\end{figure}

Therein, we have calculated the coefficients $c_n$ for every group of Fig.~\ref{fig:ewdiag}\footnote{It should be stressed that the CKM matrix is taken to be equal to the unit matrix when we calculate higher-order corrections. Consequently, Fig.~\ref{fig:ewdiag} does not contain any top quark contributions.} up to $\mathcal{O}(x^{10})$:
\begin{align}
	s^{(1)}_a &= s^{(1)}_{a1} + s^{(1)}_{a2} = -\frac{7}{6} - \frac{1}{4} \, \frac{1}{\epsilon} - \frac{1}{2} \, \mathrm{ln} \, \frac{\mu^2}{M_W^2} + \frac{1}{2} \, \mathrm{ln} \, x \notag \, ,\\
	s^{(1)}_b &= \frac{c_w}{s_w} \left( g_q - g_{q'} \right) \left[ \frac{9}{4} + \frac{3}{2} \, \frac{1}{\epsilon} + 3 \, \mathrm{ln} \, \frac{\mu^2}{M_W^2} - \frac{3}{2} \, \mathrm{ln} \, x - \frac{5}{18} \, x - \frac{5}{144} \, x^2 - \frac{23}{4200} \, x^3 - \frac{73}{75600} \, x^4 \right. \notag \\
& \qquad \qquad \qquad \qquad - \frac{53}{291060} \, x^5 - \frac{145}{4036032} \, x^6 - \frac{19}{2594592} \, x^7 - \frac{241}{157528800} \, x^8 \notag \\
& \qquad \qquad \qquad \qquad \left.  - \frac{149}{457271100} \, x^9 - \frac{19}{269549280} \, x^{10} \right] + \mathcal{O}\left(x^{11} \right) \notag \, , \\
	s^{(1)}_c &= \frac{1}{s_w^2} \, \left[ -\frac{7}{24} - \frac{1}{4} \, \frac{1}{\epsilon} - \frac{1}{2} \, \mathrm{ln} \, \frac{\mu^2}{M_W^2} + \frac{1}{4} \, \mathrm{ln} \, x \right] \notag \, , \\
\displaybreak
	s^{(1)}_d &= \left( g_q^2 + g_{q'}^2 \right) \, s_w^2 \, s^{(1)}_c \label{ewasymp} \, , \\
	s^{(1)}_e &= g_q \, g_{q'} \left[ \left( \frac{7}{12} + \frac{1}{2} \, \frac{1}{\epsilon} + \mathrm{ln} \, \frac{\mu^2}{M_W^2} - \frac{1}{2} \, \mathrm{ln} \, x \right) + \left(\frac{11}{18} - \frac{1}{3} \, \mathrm{ln} \, x \right) x + \left(-\frac{13}{144} + \frac{1}{12} \, \mathrm{ln} \, x \right) x^2 \right. \notag \\
& \qquad \qquad + \left(\frac{47}{1800} - \frac{1}{30} \, \mathrm{ln} \, x \right) x^3 + \left(-\frac{37}{3600} + \frac{1}{60} \, \mathrm{ln} \, x \right) x^4 + \left(\frac{107}{22050} - \frac{1}{105} \, \mathrm{ln} \, x \right) x^5 \notag \\
& \qquad \qquad  + \left(-\frac{73}{28224} + \frac{1}{168} \, \mathrm{ln} \, x \right) x^6 + \left(\frac{191}{127008} - \frac{1}{252} \, \mathrm{ln} \, x \right) x^7 \notag \\
& \qquad \qquad + \left(-\frac{121}{129600} + \frac{1}{360} \, \mathrm{ln} \, x \right) x^8 + \left(\frac{299}{490050} - \frac{1}{495} \, \mathrm{ln} \, x \right) x^9 \notag \\
& \qquad \qquad \left. + \left(-\frac{181}{435600} + \frac{1}{660} \, \mathrm{ln} \, x \right) x^{10} \right] + \mathcal{O}\left(x^{11} \right) \notag \, ,\\
	s^{(1)}_f &= \frac{15}{4} + \frac{3}{2} \, \frac{1}{\epsilon} + 3 \, \mathrm{ln} \, \frac{\mu^2}{M_W^2} - \frac{3}{2} \, \mathrm{ln} \, x + \left(-\frac{109}{96} + \frac{1}{2} \, \mathrm{ln} \, x \right) x + \left(-\frac{647}{1440} + \frac{1}{3} \, \mathrm{ln} \, x \right) x^2 \notag \\
& \, \quad + \left(-\frac{349}{1440} + \frac{1}{4} \, \mathrm{ln} \, x \right) x^3 + \left(-\frac{2549}{16800} + \frac{1}{5} \, \mathrm{ln} \, x \right) x^4+ \left(-\frac{419}{16800} + \frac{1}{6} \, \mathrm{ln} \, x \right) x^5  \notag \\
& \, \quad + \left(-\frac{6403}{84672} + \frac{1}{7} \, \mathrm{ln} \, x \right) x^6 + \left(-\frac{1159}{20160} + \frac{1}{8} \, \mathrm{ln} \, x \right) x^7 + \left(-\frac{1171}{25920} + \frac{1}{9} \, \mathrm{ln} \, x \right) x^8 \notag \\
& \, \quad + \left(-\frac{787}{21600} + \frac{1}{10} \, \mathrm{ln} \, x \right) x^9 + \left(-\frac{4531}{151008} + \frac{1}{11} \, \mathrm{ln} \, x \right) x^{10} + \mathcal{O}\left(x^{11} \right) \notag \, ,\\
	s^{(1)}_g &= -\frac{1}{16} \, x^2 - \frac{19}{180} \, x^3 - \frac{37}{336} \, x^4 -\frac{59}{560} \, x^5 -\frac{85}{864} \, x^6 - \frac{23}{252} \, x^7 - \frac{149}{1760} \, x^8 - \frac{17}{216} \, x^9 \notag \\
& \, \quad - \frac{229}{3120} \, x^{10} + \mathcal{O}\left(x^{11} \right) \, . \tag{\ref{ewasymp}}
\end{align}
$g_q$ and $g_{q'}$ denote the coupling of the $Z$ boson to the quarks,
\begin{equation}
\label{coupling}
	g_q = \frac{1}{s_w \, c_w} \, \left( I_q^3 - s_w^2 \, Q_q \right) \qquad \text{and} \qquad g_{q'} = \frac{1}{s_w \, c_w} \, \left( I_{q'}^3 - s_w^2 \, Q_{q'} \right) \, ,
\end{equation}
where $I_q^3 = \nicefrac{1}{2}$ ($I_{q'}^3 = \nicefrac{-1}{2}$) is the third component of the quarks' weak isospin and $Q_q = \nicefrac{2}{3}$ ($Q_{q'} = \nicefrac{-1}{3}$) is their charge. Substituting these relations into the asymptotic series (Eqs.~\eqref{ewasymp}) of the various diagrams and adding them up yields the asymptotic series of the entire electroweak contribution:
\begin{equation}
\label{ewasympall}
	S^{(1)} = \sum_{i=a}^g S^{(1)}_i \, .
\end{equation}
Note that Eqs.~\eqref{ewasymp} are specified for a special case, namely $M_Z = M_W$. In the general case, the $S$-series are accompanied by additional terms originating from the unique property of group~$(b)$: These diagrams are the only ones which contain both the $W$ and the $Z$ boson so that integrals with an additional mass scale have to be solved. In order to circumvent this problem, we suppose the mass difference of the two bosons to vanish initially. Eventually, it is accounted for by performing a Taylor expansion in the mass difference. Within Eqs.~\eqref{ewasymp}, this would lead to additional expressions of the form
\begin{equation}
	\Gamma^{(0)} \, \frac{\alpha}{\pi} \, \sum_{n=0}^{9} \, \sum_{m=1}^{10-n} \, a_{n,m} \, x^n \, \delta^m
\label{ewdelta}
\end{equation}
with
\begin{equation}
	\delta = \frac{M_W^2 - M_Z^2}{M_W^2} = -\frac{s_w^2}{c_w^2} \approx -0.29
\label{delta}
\end{equation}
in the $S$-series of every group involving a $Z$ boson. The coefficients $a_{n,m}$ are specified in \mbox{Appendix \ref{sec:ewcoeff}}.\\
Adding the $\delta^m$-terms to the $S$-series of the entire electroweak contribution for vanishing~$\delta$ ($S^{(1)}$ in Eq.~\eqref{ewasympall}) yields $S^{(1)}_\delta$, the $S$-series including $\delta$. $S^{(1)}_\delta$ can be used to examine the convergence of the Taylor expansion by studying the difference of the $S$-series including powers up to $\delta^{p+1}$ and $\delta^p$ for increasing $p$:
\begin{align}
\label{ewtaylorconv}
	\Delta_p S^{(1)}_\delta &\equiv \Gamma^{(0)} \, \frac{\alpha}{\pi} \, \sum_{n=0}^9 \, x^n \, \left( \sum_{m=1}^p a_{n,m} \, \delta^m - \sum_{m=1}^{p-1} a_{n,m} \, \delta^m \right) \notag \\
	&= \Gamma^{(0)} \, \frac{\alpha}{\pi} \, \sum_{n=0}^9 \, x^n \, a_{n,p} \, \delta^p \hspace{4cm} (p = 2..10) \, .
\end{align}
At NLO, we have calculated these coefficients up to $\mathcal{O}(\delta^{10})$ so that convergence can be examined with the help of Table~\ref{tab:ewtaylorconv}. There, we can read off that the Taylor series in the mass difference of the two bosons converges rapidly at NLO and conclude that a Taylor expansion up to $\mathcal{O}(\delta^{5})$ should be sufficient at NNLO.
\begin{table}[tb]
\centering
\caption[Convergence of the Taylor expansion in $\delta$ for the entire NLO electroweak contribution]{\textbf{Convergence of the Taylor expansion in $\boldsymbol{\delta}$ for the entire NLO electroweak contribution} expressed through $\Delta_p S^{(1)}_\delta$ as defined in Eq.~\eqref{ewtaylorconv}. All quantities are indicated for $\left| V_{qq'} \right|^2 = 1$. The nume\-rical values are given in MeV and the input para\-meters can be found in Section~\ref{sec:num}.}
\begin{tabular}{cl}
\toprule
$\Delta_2 S^{(1)}_\delta$ & $-0.017$ \\
$\Delta_3 S^{(1)}_\delta$ & $+0.011$ \\
$\Delta_4 S^{(1)}_\delta$ & $-3.69 \cdot 10^{-3}$ \\
$\Delta_5 S^{(1)}_\delta$ & $+5.04 \cdot 10^{-4}$ \\
$\Delta_6 S^{(1)}_\delta$ & $-4.63 \cdot 10^{-4}$ \\
$\Delta_7 S^{(1)}_\delta$ & $-2.77 \cdot 10^{-5}$ \\
$\Delta_8 S^{(1)}_\delta$ & $-3.54 \cdot 10^{-5}$ \\
$\Delta_9 S^{(1)}_\delta$ & $-2.26 \cdot 10^{-6}$ \\
$\Delta_{10} S^{(1)}_\delta$ & $-2.37 \cdot 10^{-7}$ \\
\midrule
$S^{(1)}_\delta$ & \multicolumn{1}{c}{$4.94$} \\
\bottomrule
\label{tab:ewtaylorconv}
\end{tabular}
\end{table}

Similarly, the set of Eqs.~\eqref{ewasymp} can be used to examine the convergence of the asymptotic expansion by studying the difference of the $S$-series including powers up to $x^{j+1}$ and $x^j$ for increasing $j$. This has been done for $\delta = 0$ since the Taylor expansion in $\delta$ has proven to converge:
\begin{align}
\label{ewasympconv}
	\Delta_j S^{(1)} &\equiv \Gamma^{(0)} \, \frac{\alpha}{\pi} \, \left( \sum_{n=0}^j c_n \, x^n - \sum_{n=0}^{j-1} c_n \, x^n \right) \notag \\
	&= \Gamma^{(0)} \, \frac{\alpha}{\pi} \, c_j \, x^j \hspace{4cm} (j = 1..10) \, .
\end{align}
\mbox{Table \ref{tab:ewasympconv}} shows this difference for each group of Fig.~\ref{fig:ewdiag} so that the convergent diagrams can be separated from the slowly converging ones. $(a)$, $(c)$ and $(d)$ obviously belong to the convergent ones since their only non-vanishing coefficients at NLO are of $\mathcal{O}(x^0)$. Groups~$(b)$ and $(e)$ converge sufficiently fast within the accuracy of the final result $S^{(1)}$. Hence, the contribution of the convergent groups~$(a)-(e)$ to the electroweak corrections is immediately obtained by approaching the on-shell value after adding the $S$-series of these diagrams:
\begin{equation}
	\label{ewwidthconv}
	\Gamma^{(1)}_{\mathrm{conv}} = \sum_{i=a}^e \Gamma^{(1)}_i = \lim_{x \to 1} \, \sum_{i=a}^e S^{(1)}_i \, .
\end{equation}
\begin{table}[tb]
\caption[Convergence of the asymptotic expansion in $x$ for the various NLO electroweak contributions]{\textbf{Convergence of the asymptotic expansion in $\boldsymbol{x}$ for the various NLO electroweak contributions} expressed through $\Delta_j S^{(1)}$ as defined in Eq.~\eqref{ewasympconv} for each group~$(a)-(g)$ of Fig.~\ref{fig:ewdiag} and their sum $\sum$. All quantities $S^{(1)}$ are indicated for $\left| V_{qq'} \right|^2 = 1$. The numerical values are given in MeV and the input parameters can be found in Section~\ref{sec:num}.}
\begin{tabularx}{1\textwidth}{cclccllll}
\toprule
 & \multicolumn{1}{c}{$(a)$} & \multicolumn{1}{c}{$(b)$} & \multicolumn{1}{c}{$(c)$} & \multicolumn{1}{c}{$(d)$} & \multicolumn{1}{c}{$(e)$} & \multicolumn{1}{c}{$(f)$} & \multicolumn{1}{c}{$(g)$} & \multicolumn{1}{c}{$\sum$} \\
\midrule
$\Delta_2 S^{(1)}$ & $0$ & $-0.19$ & $0$ & $0$ & $+0.12$ & $-0.71$ & $-0.10$ & $-0.88$ \\
$\Delta_3 S^{(1)}$ & $0$ & $-0.03$ & $0$ & $0$ & $-0.04$ & $-0.38$ & $-0.17$ & $-0.62$ \\
$\Delta_4 S^{(1)}$ & $0$ & $-5.33 \cdot 10^{-3}$ & $0$ & $0$ & $+0.01$ & $-0.24$ & $-0.17$ & $-0.41$ \\
$\Delta_5 S^{(1)}$ & $0$ & $-1.01 \cdot 10^{-3}$ & $0$ & $0$ & $-6.64 \cdot 10^{-3}$ & $-0.17$ & $-0.17$ & $-0.34$ \\
$\Delta_6 S^{(1)}$ & $0$ & $-1.98 \cdot 10^{-4}$ & $0$ & $0$ & $+3.54 \cdot 10^{-3}$ & $-0.12$ & $-0.16$ & $-0.27$ \\
$\Delta_7 S^{(1)}$ & $0$ & $-4.04 \cdot 10^{-5}$ & $0$ & $0$ & $-2.06 \cdot 10^{-3}$ & $-0.09$ & $-0.15$ & $-0.24$ \\
$\Delta_8 S^{(1)}$ & $0$ & $-8.45 \cdot 10^{-6}$ & $0$ & $0$ & $+1.28 \cdot 10^{-3}$ & $-0.07$ & $-0.14$ & $-0.20$ \\
$\Delta_9 S^{(1)}$ & $0$ & $-1.80 \cdot 10^{-6}$ & $0$ & $0$ & $-8.35 \cdot 10^{-4}$ & $-0.06$ & $-0.13$ & $-0.18$ \\
$\Delta_{10} S^{(1)}$ & $0$ & $-3.89 \cdot 10^{-7}$ & $0$ & $0$ & $+5.68 \cdot 10^{-4}$ & $-0.05$ & $-0.12$ & $-0.16$ \\
\midrule
$S^{(1)}$ & \multicolumn{1}{c}{$-1.85$} & \multicolumn{1}{c}{$10.66$} & \multicolumn{1}{c}{$-2.07$} & \multicolumn{1}{c}{$-0.81$} & \multicolumn{1}{c}{$-1.54$} & \multicolumn{1}{c}{$2.25$} & \multicolumn{1}{c}{$-1.28$} & \multicolumn{1}{c}{$5.36$} \\
\bottomrule
\label{tab:ewasympconv}
\end{tabularx}
\end{table}

\begin{figure}[tb]
	\begin{center}
	\includegraphics[scale=0.67]{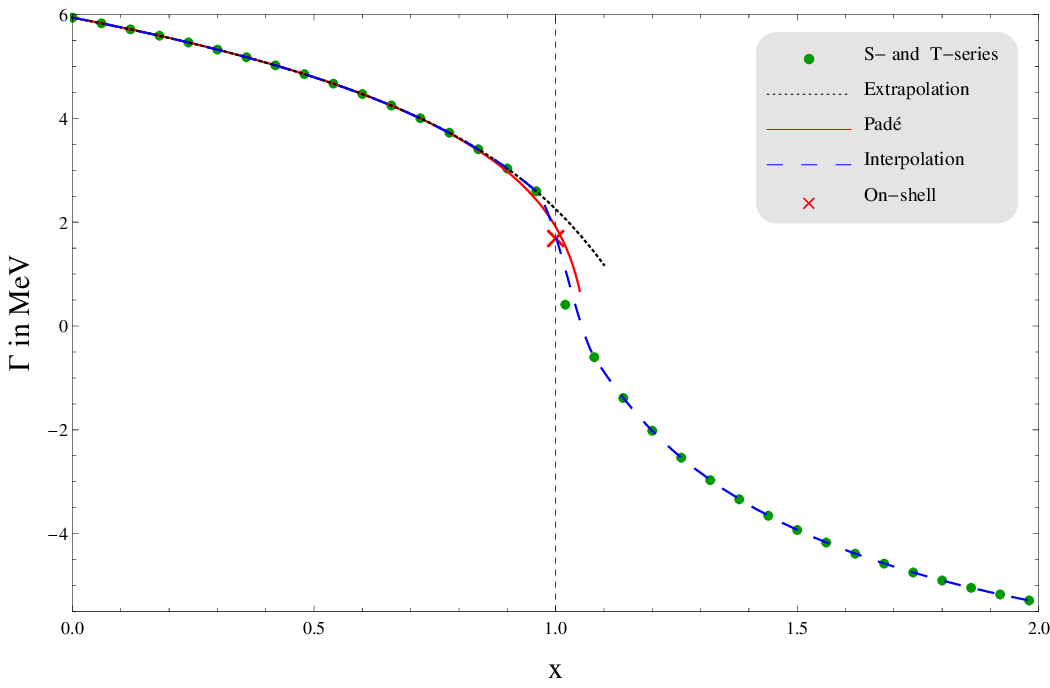}
	\caption[NLO plot for diagram $(f)$]{\textbf{NLO plot for diagram $\boldsymbol{(f)}$} including the $S$-series, the $T$-series, the extra\-polation of the $S$-series, the $[5,5]$ Pad\'{e} approximation of the $S$-series, the interpolation of the $S$- and $T$-series and the exact on-shell result. Every series is plotted up to $\mathcal{O}(x^{\pm10})$.}
	\label{fig:ewdivplot1}
	\end{center}
\end{figure}

\begin{figure}[tb]
	\begin{center}
	\includegraphics[scale=0.67]{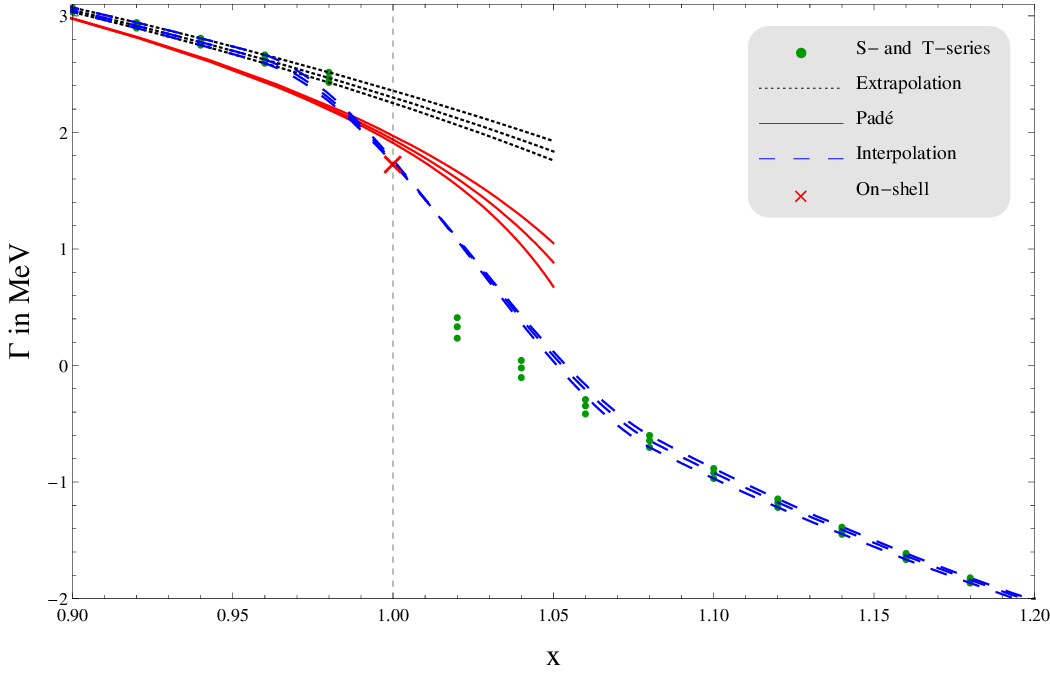}
	\caption[NLO plot for diagram $(f)$ on a larger scale]{\textbf{NLO plot for diagram $\boldsymbol{(f)}$ on a larger scale} including the $S$-series, the $T$-series, the extrapolation of the $S$-series, the Pad\'{e} approximation of the $S$-series, the interpolation of the $S$- and $T$-series and the exact on-shell result. The $S$-series and the extrapolation (the $T$-series and the interpolation on the right-hand side) are plotted up to $\mathcal{O}(x^{\pm8})$, $\mathcal{O}(x^{\pm9})$ and $\mathcal{O}(x^{\pm10})$ corresponding to the three curves from top to bottom (from bottom to top), respectively. The $[4,4]$, $[5,4]$ and $[5,5]$ Pad\'{e} approximations correspond to the three curves from top to bottom.}
	\label{fig:ewdivplot1zoom}
	\end{center}
\end{figure}

\begin{figure}[t]
	\begin{center}
	\includegraphics[scale=0.67]{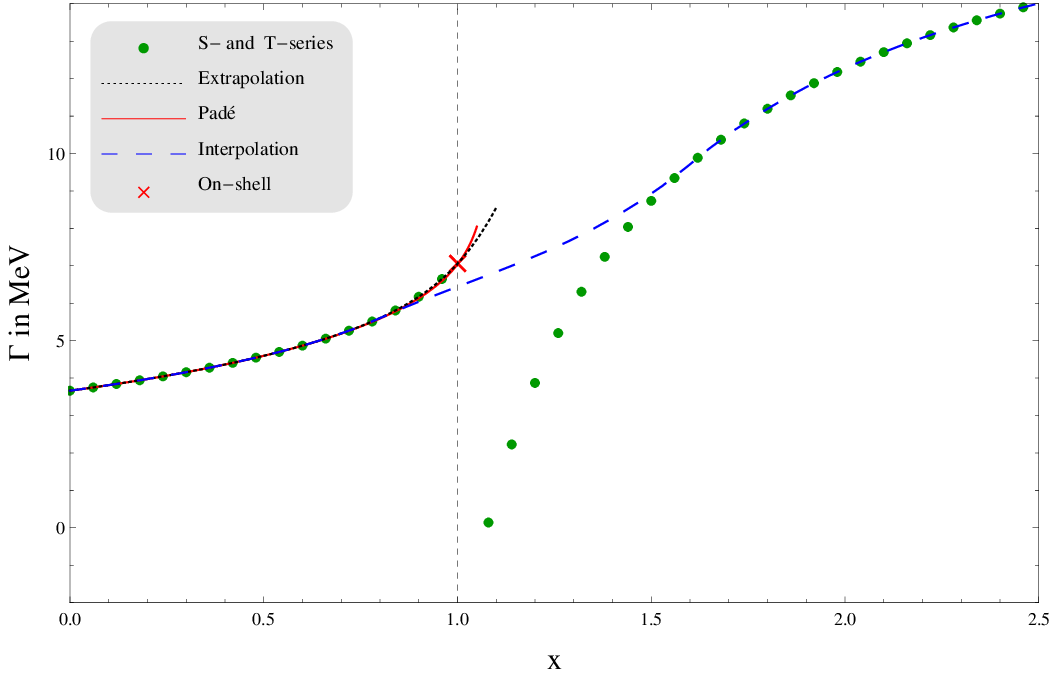}
	\caption[Combined NLO plot for the diagrams $(g)$ and $(h)$]{\textbf{Combined NLO plot for the diagrams $\boldsymbol{(g)}$ and $\boldsymbol{(h)}$} including the \mbox{$S$-series}, the $T$-series, the extrapolation of the $S$-series, the $[5,5]$ Pad\'{e} approximation of the \mbox{$S$-series}, the interpolation of the $S$- and $T$-series and the exact on-shell result. Every series is plotted up to $\mathcal{O}(x^{\pm10})$.}
	\label{fig:ewdivplot2}
	\end{center}
\end{figure}

\begin{figure}[b!]
	\begin{center}
	\includegraphics[scale=0.67]{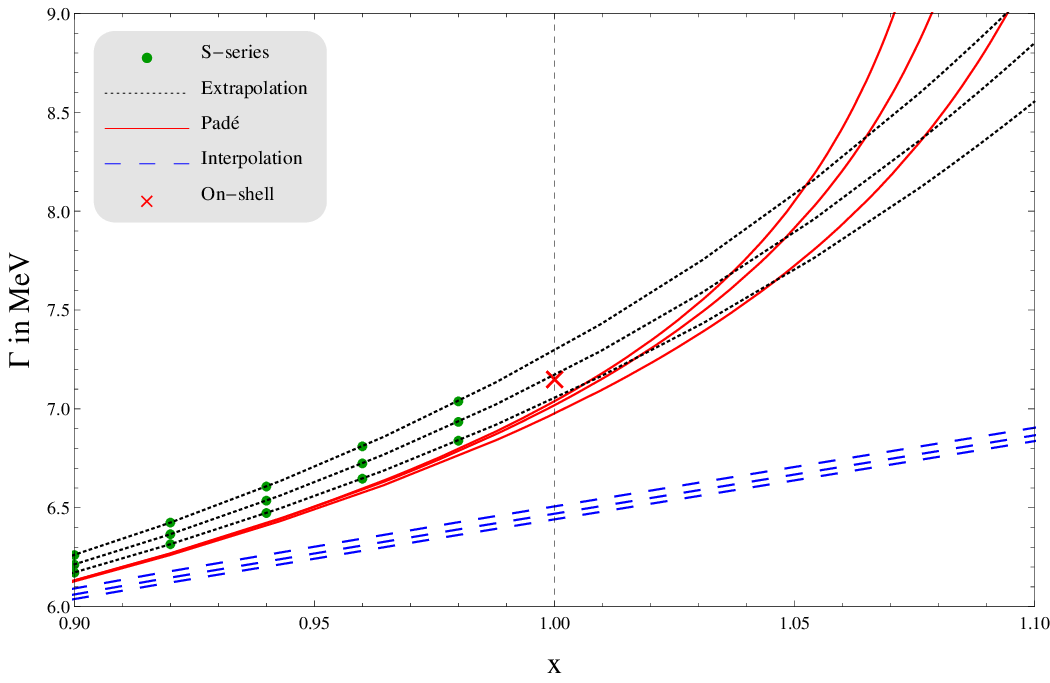}
	\caption[Combined NLO plot for the diagrams $(g)$ and $(h)$ on a larger scale]{\textbf{Combined NLO plot for the diagrams $\boldsymbol{(g)}$ and $\boldsymbol{(h)}$ on a larger scale} including the $S$-series, the extrapolation of the $S$-series, the Pad\'{e} approximation of the \mbox{$S$-series}, the interpolation of the $S$- and $T$-series and the exact on-shell result. The \mbox{$S$-series}, the extrapolation and the interpolation are plotted up to $\mathcal{O}(x^{\pm8})$, $\mathcal{O}(x^{\pm9})$ and $\mathcal{O}(x^{\pm10})$ corresponding to the three curves from top to bottom, respectively. The $[4,4]$, $[5,4]$ and $[5,5]$ Pad\'{e} approximations correspond to the three curves from bottom to top.}
	\label{fig:ewdivplot2zoom}
	\end{center}
\end{figure}

\begin{figure}[h]
	\begin{center}
	\includegraphics[scale=0.5]{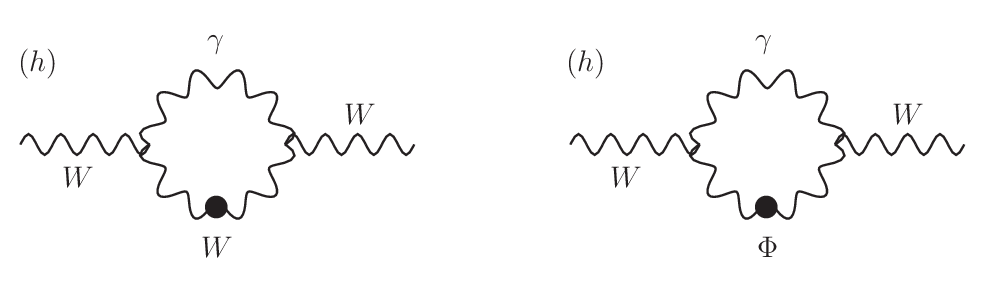}
	\caption[One-loop diagrams for the calculation of the renormalization constant $\delta Z^\mathrm{D}_W$]{\textbf{One-loop diagrams $\boldsymbol{(h)}$ for the calculation of the renormalization constant $\boldsymbol{\delta Z^\mathrm{D}_W}$} designed to cancel the on-shell IR divergence of diagram $(g)$. $\Phi$ stands for charged Goldstone bosons whereas the dots on the lines indicate that the associated propagator appears twice as a consequence of the derivative with respect to $q^2$.}
	\label{fig:deltaZW}
	\end{center}
\end{figure}

The diagrams $(f)$ and $(g)$, the only ones with a $W$ boson plus a photon inside the loop, clearly do not converge sufficiently fast, i.e. the limit of this series for $x \rightarrow 1$ does not exist within the scope of our calculations. At NLO, the method to solve the problems relating to the convergence behavior of $(f)$ and $(g)$ simply consists in computing the exact on-shell result. Unfortunately, the computation of the exact on-shell result at NNLO is much more complicated. In order to deal with this, we have developed three approximation methods at NLO, which will be applied to the slowly converging NNLO diagrams:
\begin{itemize}
\item \textbf{Extrapolation}\\
The $S$-series does not converge for $x \rightarrow 1$, but evidently it does for $x \ll 1$. Consequently, the $S$-series is extrapolated.
\item \textbf{Extrapolation $+$ Pad\'{e} approximation}\\
As for the first method, the $S$-series is extrapolated. On top of that, a Pad\'{e} approximation is performed.
\item \textbf{Interpolation}\\
We stated that the inverse asymptotic series `$T$' (`large momentum procedure', $x~\gg~1$) involves cuts of massive boson lines. Its convergence behavior turns out to be even worse than that of the $S$-series so that the limit $x \rightarrow 1$ will not exist either. Similar to the $S$-series in Eq.~\eqref{ewasympnotation}, the $T$-series is defined by
\begin{equation}
	T^{(1)} = \Gamma^{(0)} \, \frac{\alpha}{\pi} \, \sum_{n=0}^{\infty} \, \frac{\tilde{c}_n}{x^n} \equiv \Gamma^{(0)} \, \frac{\alpha}{\pi} \, t^{(1)}
\label{ewasympnotationinv}
\end{equation}
and should clearly converge for $x \gg 1$. Hence, additional information is included by approaching the threshold from the other side, too: The $T$-series is calculated and the result is approached by interpolating the $S$- and $T$-series.
\end{itemize}
In this section, we only lay the foundation for the corresponding NNLO calculation, which consists in comparing the results of the three approximation methods to the exact on-shell result at NLO. The application of these methods to the NNLO case will be discussed in Section \ref{sec:ana3}.\\
The contribution of the slowly converging diagrams to the electroweak corrections is obtained by adding the exact on-shell results of $(f)$ and $(g)$:
\begin{equation}
	\label{ewwidthdiv}
	\Gamma^{(1)}_{\mathrm{slow}} = \Gamma^{(1)}_{f} + \Gamma^{(1)}_{g} \, .
\end{equation}
Computing the exact on-shell result of $(f)$ yields
\begin{equation}
	\Gamma^{(1)}_{f} = \Gamma^{(0)} \, \frac{\alpha}{\pi} \, \left( -2 \, \zeta(2) + \frac{317}{72} + \frac{3}{2} \, \frac{1}{\epsilon} + 3 \, \mathrm{ln} \, \frac{\mu^2}{M_W^2} \right) \label{onshellf} \, ,
\end{equation}
where $\zeta(2) = \pi^2/6$. This result has to be compared to the approximated ones. The second method requires the input of the diagonal $[5,5]$ Pad\'{e} approximant of $S^{(1)}_f$:
\begin{equation}
	P^{(1)}_f [5,5] = \frac{-0.047 \, x^5 +1.046 \, x^4 -6.245 \, x^3 +15.004 \, x^2 -15.674 \, x +5.939}{-3.04 \cdot 10^{-3} \, x^5 + 0.092 \, x^4 -0.680 \, x^3 +1.939 \, x^2 -2.336 \, x +1} \, .
	\label{ewpade}
\end{equation}
The third method requires the input of the $T$-series of the diagrams $(f)$:
\begin{align}
	t^{(1)}_f &= -\frac{44}{9} + \frac{1}{6} \, \frac{1}{\epsilon} + \frac{1}{3} \, \mathrm{ln} \, \frac{\mu^2}{M_W^2} - \frac{17}{6} \, \mathrm{ln} \, x + \left(\frac{43}{24} + \frac{7}{4} \, \frac{1}{\epsilon} + \frac{7}{2} \, \mathrm{ln} \, \frac{\mu^2}{M_W^2} - \frac{51}{8} \, \mathrm{ln} \, x \right) \frac{1}{x} \notag \\
& \, \quad + \left(\frac{161}{72} - \frac{5}{12} \, \frac{1}{\epsilon} - \frac{5}{6} \, \mathrm{ln} \, \frac{\mu^2}{M_W^2} + \frac{7}{24} \, \mathrm{ln} \, x \right) \frac{1}{x^2} + \left(\frac{101}{288} - \frac{1}{3} \, \mathrm{ln} \, x \right) \frac{1}{x^3} \notag \\
& \, \quad + \left(\frac{49}{160} - \frac{1}{4} \, \mathrm{ln} \, x \right) \frac{1}{x^4} + \left(\frac{1559}{7200} - \frac{1}{5} \, \mathrm{ln} \, x \right) \frac{1}{x^5} + \left(\frac{317}{2016} - \frac{1}{6} \, \mathrm{ln} \, x \right) \frac{1}{x^6} \notag \\
& \, \quad + \left(\frac{1859}{15680} - \frac{1}{7} \, \mathrm{ln} \, x \right) \frac{1}{x^7} + \left(\frac{1117}{12096} - \frac{1}{8} \, \mathrm{ln} \, x \right) \frac{1}{x^8} + \left(\frac{13403}{181440} - \frac{1}{9} \, \mathrm{ln} \, x \right) \frac{1}{x^9} \notag \\
& \, \quad + \left(\frac{1063}{17600} - \frac{1}{10} \, \mathrm{ln} \, x \right) \frac{1}{x^{10}} + \mathcal{O}\left(\frac{1}{x^{11}} \right) \, .
\end{align}
The asymptotic expansions and the three methods as well as the on-shell result are shown in Fig.~\ref{fig:ewdivplot1}. Fig.~\ref{fig:ewdivplot1zoom} contains the same curves on a larger scale plus curves for lower-order series up to $\mathcal{O}(x^8)$ and $\mathcal{O}(x^9)$ as well as their associated Pad\'{e} approximants $P^{(1)} [5,4]$ and $P^{(1)} [4,4]$.\footnote{Alternatively, $P^{(1)} [4,5]$ may be used as Pad\'{e} approximation for the series up to $\mathcal{O}(x^9)$. For all diagrams, this approximant yields worse on-shell approximated values compared to its inverse $P^{(1)} [5,4]$. Consequently, we will use Pad\'{e} approximations of the form $P^{(1)} [n/2+1/2,n/2-1/2]$ whenever we have to apply them to series up to $\mathcal{O}(x^n)$ with odd number $n$.} Although for this diagram the interpolation method seems to agree best with the exact on-shell result at NLO, at NNLO we will use the numerical value obtained by the Pad\'{e} approximation for the following reasons:
\begin{itemize}
\item First, the Pad\'{e} approximant is better suited than the extrapolation and interpolation methods for every other diagram and we would like to find a consistent method for all diagrams.
\item Second and more important, we would like to avoid results depending on asymptotic expansions for $x \gg 1$ as far as possible. They still involve problems with respect to their convergence, which are associated with cuts of massive boson lines. Hence, the interpolation method will be only used as a cross check.
\end{itemize}
Let us proceed with diagram $(g)$. For this purpose, we introduce additional diagrams which originate from the field renormalization constant of the $W$ boson,
\begin{equation}
\delta Z_W = -\mathrm{Re} \, \frac{\partial}{\partial q^2} \, \Pi^W_T (M_W^2) \, ,
\end{equation}
where $\Pi^W_T$ is the transversal part of the $W$ boson two-point function. The derivative with respect to $q^2$ of its contributions with a photon inside the loop can be depicted via group~$(h)$ in Fig.~\ref{fig:deltaZW}. In the following, only the sum of $(g)$ and $(h)$ will be considered. This is due to infrared (IR) divergences which occur solely in the on-shell computations of $(g)$ and $(h)$. Consequently, they do not match the off-shell calculations at all and the approximation methods cannot be applied. However, both the on- and the off-shell result of the combination $(g)+(h)$ is IR finite so that Figs. \ref{fig:ewdivplot2} and \ref{fig:ewdivplot2zoom} provide a basis well suited for the determination of the NNLO result through a Pad\'{e} approximated $S$-series:
\begin{equation}
	P^{(1)}_{g+h} [5,5] = \frac{-0.010 \, x^5 -0.085 \, x^4 +0.504 \, x^3 +0.092 \, x^2 -2.001 \, x +1.573}{1.04 \cdot 10^{-3} \, x^5 + 0.036 \, x^4 -0.459 \, x^3 +1.611 \, x^2 -2.174 \, x +1} \, .
\end{equation}
\clearpage
\clearpage
The individual contributions of group $(g)$ to the exact on-shell result and the $t$-series are given by
\begin{align}
		\Gamma^{(1)}_{g} &= \Gamma^{(0)} \, \frac{\alpha}{\pi} \, \left( \frac{29}{9} + \frac{1}{2} \, \frac{1}{\epsilon_{\mathrm{IR}}} + \mathrm{ln} \, \frac{\mu^2}{M_W^2} \right)
\label{onshellg} \, , \\
	t^{(1)}_g &= \frac{689}{72} + \frac{11}{12} \, \frac{1}{\epsilon} + \frac{11}{6} \, \mathrm{ln} \, \frac{\mu^2}{M_W^2} - \frac{11}{6} \, \mathrm{ln} \, x + \left(-\frac{169}{24} - \frac{11}{4} \, \frac{1}{\epsilon} - \frac{11}{2} \, \mathrm{ln} \, \frac{\mu^2}{M_W^2} + \frac{11}{2} \, \mathrm{ln} \, x \right) \frac{1}{x} \notag \\
& \, \quad + \left(-\frac{361}{72} + \frac{5}{6} \, \frac{1}{\epsilon} + \frac{5}{3} \, \mathrm{ln} \, \frac{\mu^2}{M_W^2} - \frac{5}{3} \, \mathrm{ln} \, x \right) \frac{1}{x^2} -\frac{15}{16} \frac{1}{x^3} - \frac{19}{20} \, \frac{1}{x^4} - \frac{37}{48} \, \frac{1}{x^5} - \frac{177}{288} \, \frac{1}{x^6} \notag \\
& \, \quad  - \frac{17}{32} \, \frac{1}{x^7} - \frac{115}{252} \, \frac{1}{x^8} - \frac{447}{1220} \, \frac{1}{x^9} - \frac{17}{48} \, \frac{1}{x^{10}} + \mathcal{O}\left(\frac{1}{x^{11}}\right) \, .
\end{align}
For group $(h)$, the individual contributions read as follows:
\begin{align}
	\Gamma^{(1)}_{h} &= \Gamma^{(0)}_\epsilon \, \frac{\alpha}{\pi} \, \left( \frac{131}{72} + \frac{19}{24} \, \frac{1}{\epsilon} + \frac{19}{24} \, \mathrm{ln} \, \frac{\mu^2}{M_W^2} + \left[ -1 - \frac{1}{2} \, \frac{1}{\epsilon} - \frac{1}{2} \, \mathrm{ln} \, \frac{\mu^2}{M_W^2} \right]_{\mathrm{IR}} \right) \label{dZWgamma} \, , \\
	S^{(1)}_h &= \Gamma^{(0)}_\epsilon \, \frac{\alpha}{\pi} \, \left(\frac{143}{144} +\frac{19}{24} \, \frac{1}{\epsilon} +\frac{43}{48} \, x +\frac{15}{32} \, x^2 +\frac{23}{72} \, x^3 + \frac{163}{672} \, x^4 +\frac{219}{1120} \, x^5 \right. +\frac{283}{1728} \, x^6 \notag \\
& \qquad \qquad \quad \left. +\frac{71}{504} \, x^7 +\frac{87}{704} \, x^8 +\frac{523}{4752} \, x^9 +\frac{619}{6240} \, x^{10} \right) + \mathcal{O}\left(x^{11}\right) \label{dZWgammaS} \, ,\\
	T^{(1)}_h &= \Gamma^{(0)}_\epsilon \, \frac{\alpha}{\pi} \, \left(\frac{43}{36} + \frac{19}{24} \, \frac{1}{\epsilon} + \frac{19}{24} \, \mathrm{ln} \, \frac{\mu^2}{M_W^2} - \frac{19}{24} \, \mathrm{ln} \, x + \left(\frac{13}{16} -\frac{5}{8} \, \mathrm{ln} \, x \right)  \frac{1}{x} \right. \notag \\
& \qquad \qquad \quad + \left(\frac{8}{9} +\frac{5}{12} \, \mathrm{ln} \, x \right) \frac{1}{x^2} +\frac{3}{32} \, \frac{1}{x^3} +\frac{19}{120} \, \frac{1}{x^4} +\frac{43}{288} \, \frac{1}{x^5} +\frac{15}{112} \, \frac{1}{x^6} \notag \\
& \qquad \qquad \quad \left. +\frac{23}{192} \, \frac{1}{x^7} +\frac{163}{1512} \, \frac{1}{x^8} +\frac{219}{2240} \, \frac{1}{x^9} +\frac{283}{3168} \, \frac{1}{x^{10}} \right) + \mathcal{O}\left(\frac{1}{x^{11}}\right) \label{dZWgammaT} \, .
\end{align}
The subscript IR refers to an IR pole in contrast to a UV one (without a subscript).\\
The total contribution to the electroweak corrections is then given by the sum
\begin{equation}
	\label{ewwidth}
	\Gamma^{(1)}_{\mathrm{EW}} = \Gamma^{(1)}_{\mathrm{conv}} + \Gamma^{(1)}_{\mathrm{slow}} + \Gamma^{(1)}_{\mathrm{ren}} \, .
\end{equation}
Note that renormalization constants have to be multiplied by $\Gamma^{(0)}_\epsilon$, the Born width up to $\mathcal{O}(\epsilon)$ in Eq.~\eqref{borneps}, instead of $\Gamma^{(0)}$. That way, an additional finite part is picked up from the multiplication of the pole part of the renormalization constant by the $\mathcal{O}(\epsilon)$ term of the Born decay width. We have applied an on-shell renormalization scheme described in Ref.~\cite{Denner:1993}, leading to
\begin{equation}
	\label{Denner}
	\Gamma^{(1)}_\mathrm{ren} = 2 \, \Gamma^{(0)}_\epsilon \, \delta Z \, .
\end{equation}
\clearpage
From the renormalized charged current vertex, it follows that
\begin{align}
	\delta Z &\equiv \delta Z_e - \frac{\delta s_w}{s_w} + \frac{1}{2} \, \delta Z_W \label{Denner3} \\
	&= \frac{1}{2} \, \frac{\partial}{\partial q^2} \, \Pi^\gamma_T(0) - \frac{s_w}{c_w} \, \frac{\Pi^{\gamma Z}_T(0)}{M_Z^2} +\frac{c_w^2}{2 \, s_w^2} \, \mathrm{Re} \, \left(\frac{\Pi^W_T(M_W^2)}{M_W^2} - \frac{\Pi^Z_T(M_Z^2)}{M_Z^2} \right) -\frac{1}{2} \, \frac{\partial}{\partial q^2} \, \Pi^W_T(M_W^2) \, . \notag
\end{align}
At this point, we should comment on the parameterization of our calculations. All preceding formulae have used $\alpha$ and the physical particle masses as basic parameters. In such on-shell renormalization schemes, large electroweak corrections arise from fermion loop contributions to the renormalization of $\alpha$ and $s_w$. As in any charged-current process, these corrections can be reduced by parameterizing the lowest-order result with Fermi's coupling constant $G_F$ and $M_W$ instead \cite{Sirlin:1980}. This can be achieved with the help of the relationship
\begin{equation}
	G_F = \frac{\pi \, \alpha}{\sqrt{2} \, s_w^2 \, M_W^2} \frac{1}{1 - \Delta r} \, .
\end{equation}
$\Delta r$ contains the radiative corrections to the muon decay width which the Standard Model introduces in addition to the purely photonic corrections from within Fermi's model. At one loop, this expression is finite and given by \cite{Denner:1993,Sirlin:1980}
\begin{align}
	\Delta r &= \frac{\Pi^W_T(0) - \mathrm{Re} \, \Pi^W_T(M_W^2)}{M_W^2} + \frac{c_w^2}{s_w^2} \, \mathrm{Re} \, \left(\frac{\Pi^W_T(M_W^2)}{M_W^2} - \frac{\Pi^Z_T(M_Z^2)}{M_Z^2} \right) + 2 \, \frac{c_w}{s_w} \, \frac{\Pi^{\gamma Z}_T(0)}{M_Z^2} \notag \\
& \, \quad + \frac{\partial}{\partial q^2} \, \Pi^\gamma_T(0) + \frac{\alpha}{4 \, \pi \, s_w^2} \, \left[\left(\frac{7}{2 \, s_w^2} - 2 \right) \, \mathrm{ln} \, c_w^2 + 6 \right] \, .
\end{align}
The term in the square brackets is obtained by the vertex and box corrections to the muon decay width in Feynman-'t Hooft gauge. Hence, the self-energies $\Pi_T$ have to be calculated in the same gauge. This requires the computation of 
\begin{equation}
	\frac{\partial}{\partial q^2} \, \Pi^\gamma_T(0) \, ,
\end{equation}
which receives important contributions from the light quark flavors. They cannot be reliably predicted in perturbative QCD so that the gauge-independent and finite quantity
\begin{equation}
	\Delta \alpha^{(5)}_{\mathrm{had}} = \left[ \frac{\partial}{\partial q^2} \, \Pi^\gamma_T(0) - \frac{\Pi^\gamma_T(M_Z^2)}{M_Z^2} \right]_{udscb}
\label{delalphahad}
\end{equation}
is introduced. That way, experimental data on the total cross section of inclusive hadron production in $e^+ e^-$ annihilation can be used to circumvent the problem. However, we substitute
\begin{equation}
	\alpha = \frac{\sqrt{2} \, G_F \, s_w^2 \, M_W^2}{\pi}
\label{alphaGF}
\end{equation}
whenever the lowest-order result $\Gamma^{(0)}$ of Eq.~\eqref{born} appears within our calculations. In turn, we add the term $- \Gamma^{(0)}_\epsilon \cdot \Delta r$ to Eq.~\eqref{Denner} so that the quantity $\partial/\partial q^2 \, \Pi^\gamma_T(0)$ exactly cancels and the theoretical uncertainty of $\Delta \alpha^{(5)}_{\mathrm{had}}$ does not affect our results:
\begin{equation}
	\Gamma^{(1)}_\mathrm{ren} = \Gamma^{(0)}_\epsilon \, \left(2 \, \delta Z - \Delta r \right) \, .
\end{equation}
The pole part of $\Gamma^{(1)}_\mathrm{ren}$ in Eq.~\eqref{ewwidth} is then 	given by
\begin{equation}
	\tilde{\Gamma}^{(1)}_\mathrm{ren} = -\Gamma^{(0)} \, \frac{\alpha}{\pi} \, \left(\frac{1}{s_w^2 \, \epsilon} + \frac{1}{2} \, \frac{1}{\epsilon_{\mathrm{IR}}} \right) = -\left(\tilde{\Gamma}^{(1)}_\mathrm{conv} + \tilde{\Gamma}^{(1)}_\mathrm{slow} \right) \, .
\end{equation}
Hence, the renormalization procedure cancels both the UV and the on-shell IR divergences and thus produces a physical quantity.

\subsection{Next-To-Next-To-Leading-Order Decay Width:\texorpdfstring{\\}{} Mixed QCD/Electroweak Corrections of Order \texorpdfstring{$\boldsymbol{\alpha \, \alpha_s}$}{alpha alphas}}
\label{sec:ana3}

\begin{figure}[b]
	\begin{center}
	\includegraphics[scale=0.5]{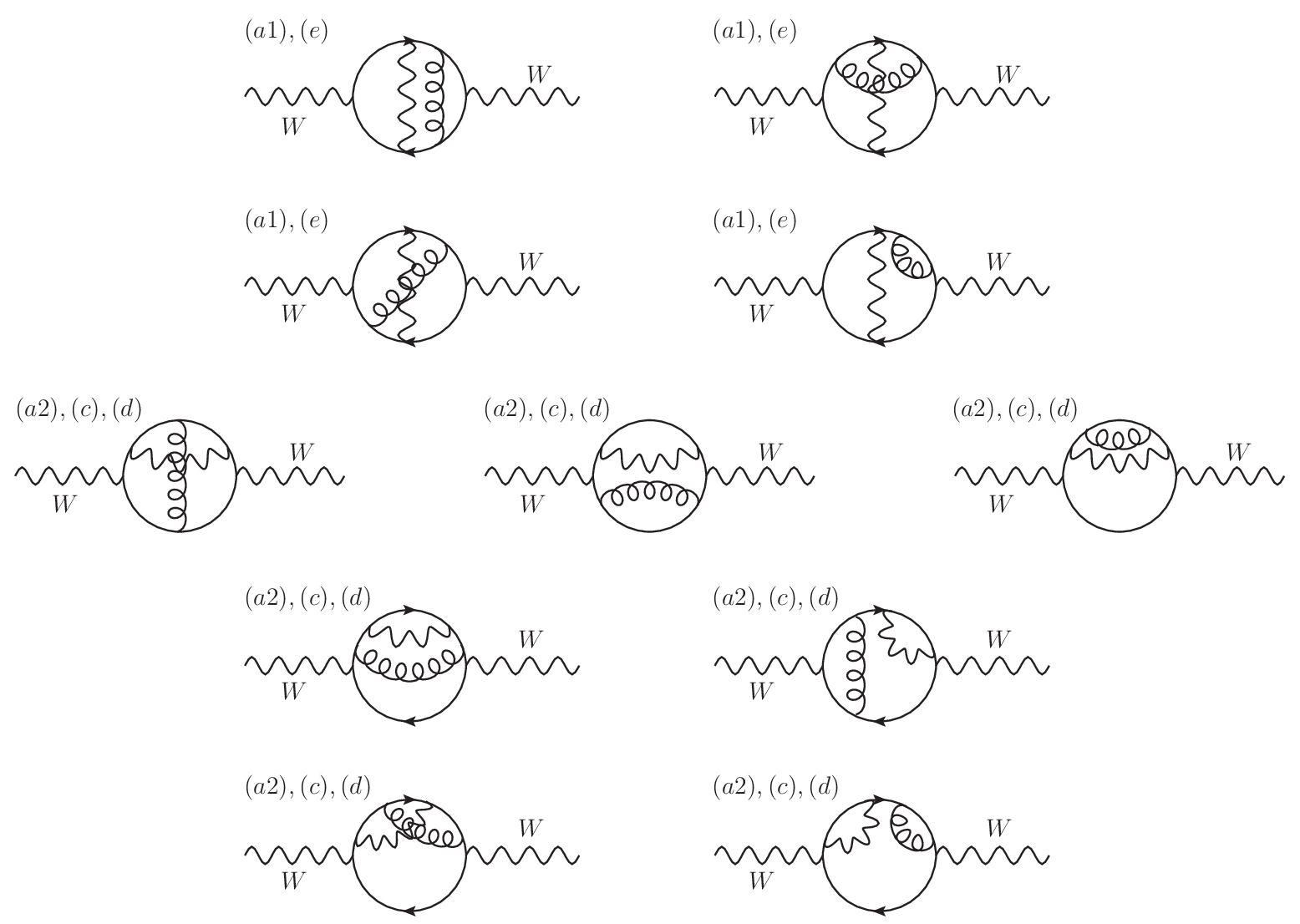}
	\includegraphics[scale=0.5]{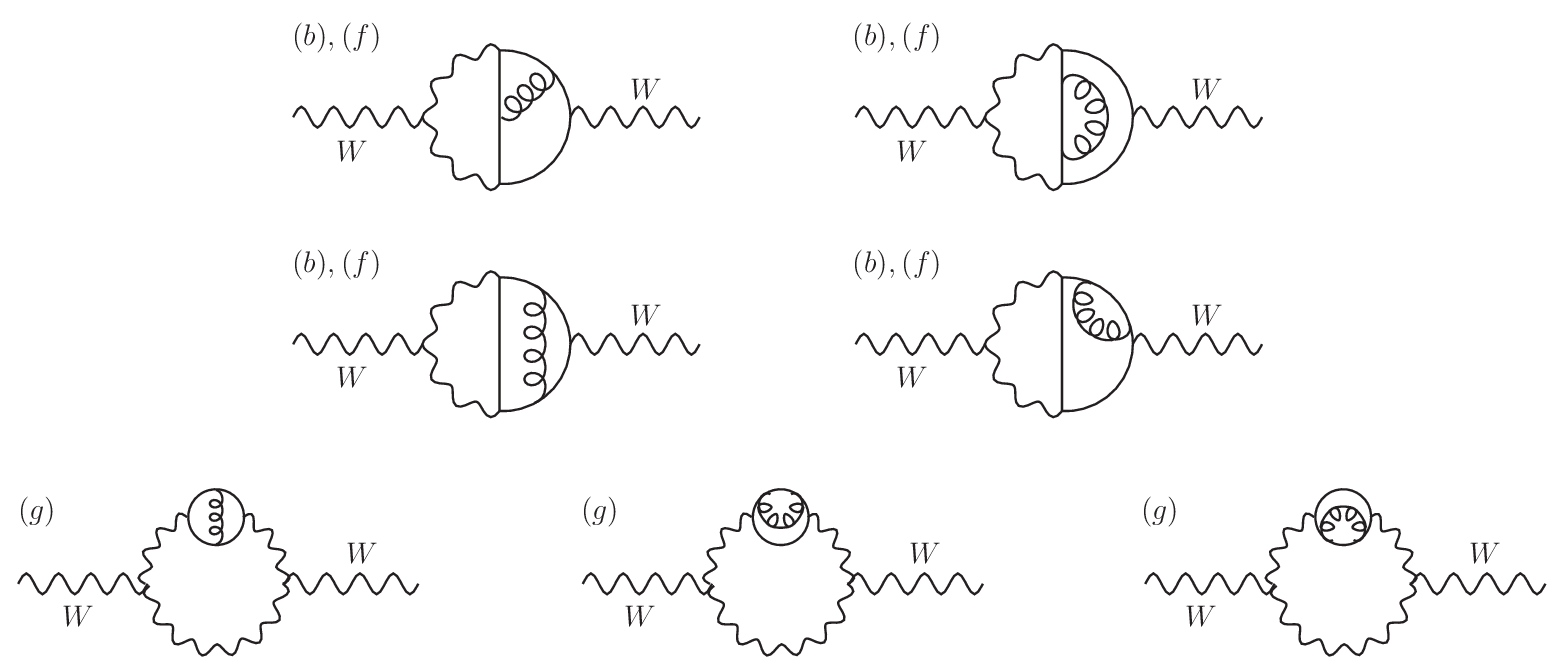}
	\caption[Three-loop diagrams for the calculation of the $\mathcal{O}(\alpha \, \alpha_s)$ mixed QCD/electroweak corrections]{\textbf{Three-loop diagrams for the calculation of the $\boldsymbol{\mathcal{O}(\alpha \, \alpha_s)}$ mixed QCD/ electroweak corrections} $\Gamma^{(2)}_\mathrm{mixed}$. The labeling corresponds to the seven groups $(a)-(g)$ of Fig. \ref{fig:ewdiag}. By adding one gluon in every possible way, these eighteen topologies occur. The straight lines stand for quarks, the curly lines for gluons and the wavy ones denote photons, $Z$ or $W$ bosons.}
	\label{fig:mixeddiag}
	\end{center}
\end{figure}

The calculation of the mixed QCD/electroweak corrections follows the same logic as the computation of the electroweak corrections in Section \ref{sec:ana2}:
\begin{equation}
	\Gamma^{(2)}_\mathrm{mixed} = \Gamma^{(2)}_\mathrm{conv} + \Gamma^{(2)}_\mathrm{slow} + \Gamma^{(2)}_\mathrm{ren} \, .
\label{mixedwidth}
\end{equation}
All diagrams in question can be obtained by adding exactly one gluon to the NLO electroweak diagrams in every possible way. Thus, we can reuse the labeling of the diagrams in Fig.~\ref{fig:ewdiag}, which is depicted in Fig.~\ref{fig:mixeddiag}.
As a first step, the convergent diagrams have to be separated from the slowly converging ones by applying the hard mass procedure. This yields a result of the form
\begin{equation}
	S^{(2)} = \Gamma^{(0)} \, \frac{\alpha}{\pi} \, \frac{\alpha_s}{\pi} \, \sum_{n=0}^\infty d_n \, x^n \equiv \Gamma^{(0)} \, \frac{\alpha}{\pi} \, \frac{\alpha_s}{\pi} \, s^{(2)} \, ,
\label{mixedasympnotation}
\end{equation}
where we have made use of the relationship
\begin{equation}
	n_c^2 = 2 \, n_c \, C_F + 1
\end{equation}
in order to factor out the Born decay width $\Gamma^{(0)}$. $C_F$ is the quadratic Casimir operator of the fundamental representation $SU(n_c)$ of the strong interaction. For three colors, it is given by $C_F = \nicefrac{4}{3}$.\\
We computed the coefficients $d_n$ for every group~$(a)-(g)$ up to $\mathcal{O}(x^{9})$:
\vspace{0.5cm}
\begin{align}
	s^{(2)}_a &= s^{(2)}_{a1} + s^{(2)}_{a2} = -\frac{433}{144} + 2 \, \zeta(3) - \frac{1}{8} \, \frac{1}{\epsilon} - \frac{3}{8} \, \mathrm{ln} \, \frac{\mu^2}{M_W^2} + \frac{3}{8} \, \mathrm{ln} \, x \notag \, , \\ & \notag \\
	s^{(2)}_b &= \frac{c_w}{s_w} \left( g_q - g_{q'} \right) \left[ \frac{103}{8} - 12 \, \zeta(3) + \frac{5}{4} \, \frac{1}{\epsilon} + \frac{15}{4} \, \mathrm{ln} \, \frac{\mu^2}{M_W^2} - \frac{11}{4} \, \mathrm{ln} \, x - \frac{7}{18} \, x - \frac{89}{2160} \, x^2 \right. \notag \\
& \qquad \qquad \qquad \qquad - \frac{29}{4725} \, x^3 - \frac{53}{50400} \, x^4 - \frac{851}{4365900} \, x^5 - \frac{1381}{36324288} \, x^6 - \frac{349}{45405360} \, x^7 \notag \\
& \qquad \qquad \qquad \qquad \left. - \frac{3017}{1890345600} \, x^8 - \frac{4177}{12346319700} \, x^9 \right] + \mathcal{O}\left(x^{10}\right) \notag \, ,\\ & \notag \\
	s^{(2)}_c &= \frac{1}{s_w^2} \, \left[ \left( -\frac{89}{48} + 2 \, \zeta(3) - \frac{1}{8} \, \frac{1}{\epsilon} - \frac{3}{8} \, \mathrm{ln} \, \frac{\mu^2}{M_W^2} + \frac{3}{8} \, \mathrm{ln} \, x \right) + \left(-\frac{121}{243} + \frac{22}{81} \, \mathrm{ln} \, x - \frac{1}{27} \, \mathrm{ln}^2 \, x\right) x \right. \notag \\
& \quad \qquad + \left(-\frac{169}{15552} + \frac{13}{1296} \, \mathrm{ln} \, x - \frac{1}{432} \, \mathrm{ln}^2 \, x\right) x^2 \notag \\
& \quad \qquad + \left(-\frac{2209}{2430000} + \frac{47}{40500} \, \mathrm{ln} \, x - \frac{1}{2700} \, \mathrm{ln}^2 \, x\right) x^3 \notag \\
& \quad \qquad + \left(-\frac{1369}{9720000} + \frac{37}{162000} \, \mathrm{ln} \, x - \frac{1}{10800} \, \mathrm{ln}^2 \, x\right) x^4 \notag \\
& \quad \qquad + \left(-\frac{11449}{364651875} + \frac{214}{3472875} \, \mathrm{ln} \, x - \frac{1}{33075} \, \mathrm{ln}^2 \, x\right) x^5 \notag \\
& \quad \qquad  + \left(-\frac{5329}{597445632} + \frac{73}{3556224} \, \mathrm{ln} \, x - \frac{1}{84672} \, \mathrm{ln}^2 \, x\right) x^6 \notag \\
& \quad \qquad + \left(-\frac{36481}{12098274048} + \frac{191}{24004512} \, \mathrm{ln} \, x - \frac{1}{190512} \, \mathrm{ln}^2 \, x\right) x^7 \notag \\
& \quad \qquad + \left(-\frac{14641}{12597120000} + \frac{121}{34992000} \, \mathrm{ln} \, x - \frac{1}{388800} \, \mathrm{ln}^2 \, x\right) x^8 \notag \\
& \quad \qquad \left. + \left(-\frac{89401}{180111751875} + \frac{598}{363862125} \, \mathrm{ln} \, x - \frac{1}{735075} \, \mathrm{ln}^2 \, x\right) x^9 \right] + \mathcal{O}\left(x^{10}\right) \, , \notag \\
\displaybreak
	s^{(2)}_d &= \left( g_q^2 + g_{q'}^2 \right) \, s_w^2 \, s^{(2)}_c \label{mixedasymp} \, , \\
	s^{(2)}_e &= g_q \, g_{q'} \left[ \left( \frac{89}{24} - 4 \, \zeta(3) + \frac{1}{4} \, \frac{1}{\epsilon} + \frac{3}{4} \, \mathrm{ln} \, \frac{\mu^2}{M_W^2} - \frac{3}{4} \, \mathrm{ln} \, x \right)  \right. \notag \\
& \qquad \qquad + \left(\frac{71}{27} - \frac{4}{3} \, \zeta(3) - \frac{2}{3} \, \mathrm{ln} \, x - \frac{1}{3} \, \mathrm{ln}^2 \, x\right) x \notag \\
& \qquad \qquad + \left(-\frac{1159}{2592} + \frac{1}{3} \, \zeta(3) + \frac{5}{6} \, \mathrm{ln} \, x - \frac{7}{36} \, \mathrm{ln}^2 \, x\right) x^2 \notag \\
& \qquad \qquad + \left(-\frac{1853}{10800} - \frac{2}{15} \, \zeta(3) - \frac{13}{90} \, \mathrm{ln} \, x \right) x^3 \notag \\
& \qquad \qquad + \left(\frac{146179}{972000} + \frac{1}{15} \, \zeta(3) + \frac{83}{810} \, \mathrm{ln} \, x - \frac{11}{180} \, \mathrm{ln}^2 \, x \right) x^4 \notag \\
& \qquad \qquad + \left(-\frac{78941}{595350} - \frac{4}{105} \, \zeta(3) - \frac{37}{2835} \, \mathrm{ln} \, x \right) x^5 \notag \\
& \qquad \qquad  + \left(\frac{514328497}{5334336000} + \frac{1}{42} \, \zeta(3) + \frac{221}{12960} \, \mathrm{ln} \, x - \frac{143}{5040} \, \mathrm{ln}^2 \, x \right) x^6 \notag \\
& \qquad \qquad + \left(-\frac{262758413}{3429216000} - \frac{1}{63} \, \zeta(3) + \frac{799}{136080} \, \mathrm{ln} \, x - \frac{1}{1080} \, \mathrm{ln}^2 \, x \right) x^7 \notag \\
& \qquad \qquad + \left(\frac{13907067061}{240045120000} + \frac{1}{90} \, \zeta(3) + \frac{853}{6804000} \, \mathrm{ln} \, x - \frac{403}{25200} \, \mathrm{ln}^2 \, x \right) x^8 \notag \\
& \qquad \qquad \left. + \left(-\frac{42384988309}{907670610000} - \frac{4}{495} \, \zeta(3) + \frac{19058}{2338875} \, \mathrm{ln} \, x - \frac{11}{9450} \, \mathrm{ln}^2 \, x \right) x^9 \right] + \mathcal{O}\left(x^{10}\right) \notag \, ,\\
	s^{(2)}_f &= \frac{111}{8} + \frac{5}{4} \, \frac{1}{\epsilon} - 12 \, \zeta(3) + \frac{15}{4} \, \mathrm{ln} \, \frac{\mu^2}{M_W^2} - \frac{11}{4} \, \mathrm{ln} \, x + \left(-\frac{1157}{864} + \frac{5}{9} \, \mathrm{ln} \, x \right) x  \notag \\
& \, \quad + \left(-\frac{2071}{4320} + \frac{25}{72} \, \mathrm{ln} \, x \right) x^2 + \left(-\frac{5423}{21600} + \frac{23}{90} \, \mathrm{ln} \, x \right) x^3 + \left(-\frac{23459}{151200} + \frac{73}{360} \, \mathrm{ln} \, x \right) x^4  \notag \\
& \, \quad + \left(-\frac{223399}{2116800} + \frac{53}{315} \, \mathrm{ln} \, x \right) x^5 + \left(-\frac{1619}{21168} + \frac{145}{1008} \, \mathrm{ln} \, x \right) x^6 \notag \\
& \, \quad + \left(-\frac{220961}{3810240} + \frac{95}{756} \, \mathrm{ln} \, x \right) x^7 + \left(-\frac{8843}{194400} + \frac{241}{2160} \, \mathrm{ln} \, x \right) x^8 \notag \\
& \, \quad + \left(-\frac{861827}{23522400} + \frac{149}{1485} \, \mathrm{ln} \, x \right) x^9 + \mathcal{O}\left(x^{10}\right) \notag \, ,\\
	s^{(2)}_g &= -\frac{1}{16} \, x^2 - \frac{19}{180} \, x^3 - \frac{37}{336} \, x^4 -\frac{59}{560} \, x^5 -\frac{85}{864} \, x^6 - \frac{23}{252} \, x^7 - \frac{149}{1760} \, x^8 - \frac{17}{216} \, x^9 \notag \\
& \, \quad + \mathcal{O}\left(x^{10}\right) \, . \tag{\ref{mixedasymp}}
\end{align}
The asymptotic series of the entire NNLO contribution is given by the sum of these expressions:
\begin{equation}
\label{mixedasympall}
	S^{(2)} = \sum_{i=a}^g S^{(2)}_i \, .
\end{equation}
As in case of the NLO electroweak analog (Eqs.~\eqref{ewasymp}), Eqs.~\eqref{mixedasymp} do not indicate the general result, but for the special case $M_Z = M_W$. Again, we account for the mass difference of the $W$ and the $Z$ boson by performing a Taylor expansion in the exact same manner as in Section \ref{sec:ana2}. In doing so, we pick up additional expressions of the form
\begin{equation}
	\label{mixeddelta}
	\Gamma^{(0)} \, \frac{\alpha}{\pi} \, \frac{\alpha_s}{\pi} \,\sum_{n=0}^{8} \, \sum_{m=1}^{\mathrm{min}(5,9-n)} b_{n,m} \, x^n \, \delta^m
\end{equation}
with
\begin{equation}
	\delta = \frac{M_W^2 - M_Z^2}{M_W^2} = -\frac{s_w^2}{c_w^2} \approx -0.29
\tag{\ref{delta}}
\end{equation}
within the $S$-series of every group involving a $Z$ boson. The coefficients $b_{n,m}$ can be found in Appendix~\ref{sec:mixedcoeff}. The upper bound of the second summation is given by the smaller number of $5$ and $9-n$, which is denoted by the function `$\mathrm{min}$'. This is due to Section~\ref{sec:ana2}, where we have decided that a Taylor expansion up to $\mathcal{O}(\delta^{5})$ should be sufficient at NNLO because of the excellent convergence behavior of the Taylor expansion at NLO.\\
By adding the $\delta^m$-terms to the $S$-series of the entire NNLO contribution for vanishing $\delta$ ($S^{(2)}$ in Eq.~\eqref{mixedasympall}), we obtain $S^{(2)}_\delta$, which is the $S$-series including $\delta$. $S^{(2)}_\delta$ can be used to study the convergence of the Taylor expansion by examining the difference of the $S$-series including powers up to $\delta^{p+1}$ and $\delta^p$ for increasing $p$:
\begin{align}
\label{mixedtaylorconv}
	\Delta_p S^{(2)}_\delta &\equiv \Gamma^{(0)} \, \frac{\alpha}{\pi} \, \frac{\alpha_s}{\pi} \,\sum_{n=0}^{8} \, x^n \, \left( \sum_{m=1}^p b_{n,m} \, \delta^m - \sum_{m=1}^{p-1} b_{n,m} \, \delta^m \right) \notag \\
	&= \Gamma^{(0)} \, \frac{\alpha}{\pi} \, \frac{\alpha_s}{\pi} \,\sum_{n=0}^{8} \, x^n \, b_{n,p} \, \delta^p \hspace{4cm} (p = 2..5) \, .
\end{align}
\begin{table}[b]
\centering
\caption[Convergence of the Taylor expansion in $\delta$ for the entire NNLO contribution]{\textbf{Convergence of the Taylor expansion in $\boldsymbol{\delta}$ for the entire NNLO contribution} expressed through $\Delta_p S^{(2)}_\delta$ as defined in Eq.~\eqref{mixedtaylorconv}. All quantities are indicated for $\left| V_{qq'} \right|^2 = 1$. The numerical values are given in keV and the input parameters can be found in Section~\ref{sec:num}.}
\begin{tabular}{cl}
\toprule
$\Delta_2 S^{(2)}_\delta$ & $+2.8$ \\
$\Delta_3 S^{(2)}_\delta$ & $+5.3$ \\
$\Delta_4 S^{(2)}_\delta$ & $+2.5$ \\
$\Delta_5 S^{(2)}_\delta$ & $+1.0$ \\
\midrule
$S^{(2)}_\delta$ & \multicolumn{1}{c}{$-642.0$} \\
\bottomrule
\label{tab:mixedtaylorconv}
\end{tabular}
\end{table}

\begin{figure}[tb]
	\begin{center}
	\includegraphics[scale=0.67]{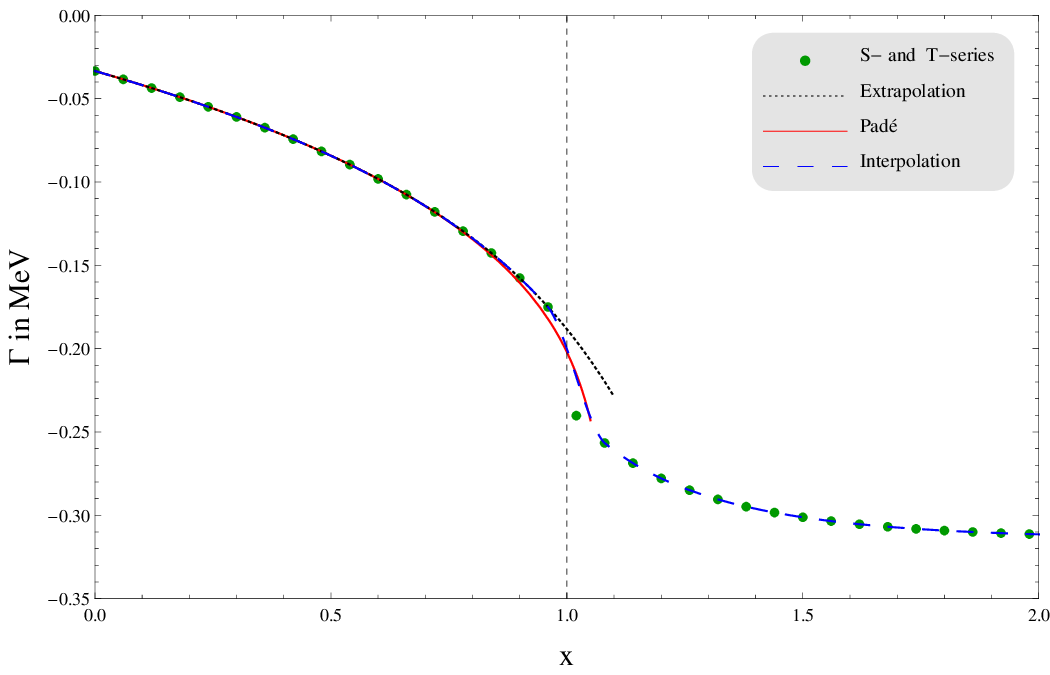}
	\caption[NNLO plot for diagram $(f)$]{\textbf{NNLO plot for diagram $\boldsymbol{(f)}$} including the $S$-series, the $T$-series, the extra\-polation of the $S$-series, the $[5,4]$ Pad\'{e} approximation of the $S$-series and the interpolation of the $S$- and $T$-series. Every series is plotted up to $\mathcal{O}(x^{\pm9})$.}
	\label{fig:mixeddivplot1}
	\end{center}
\end{figure}

\begin{figure}[tb]
	\begin{center}
	\includegraphics[scale=0.67]{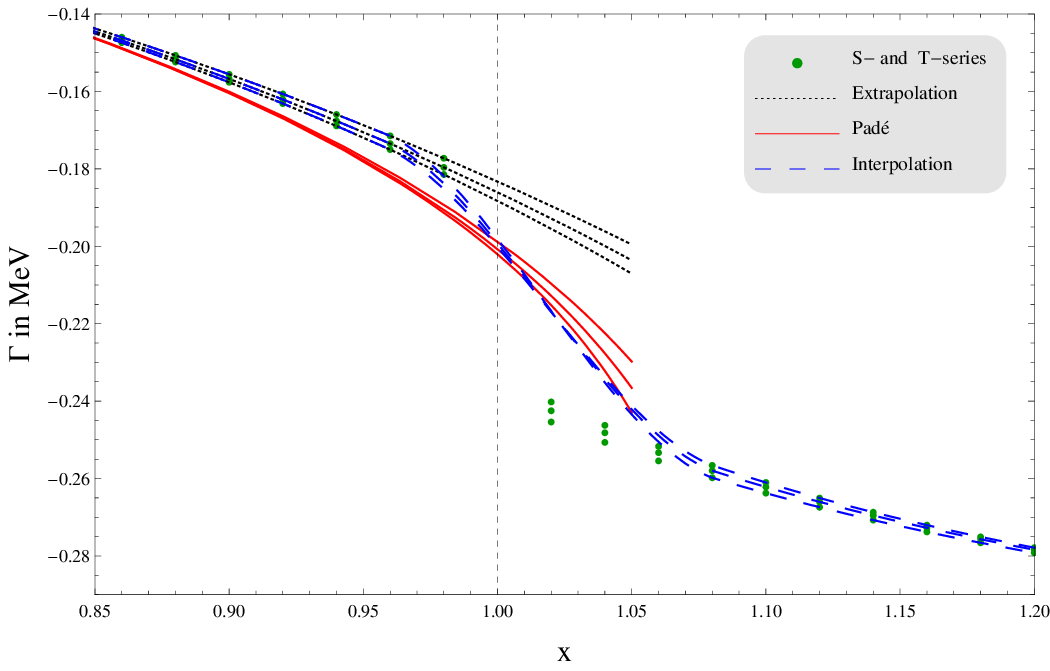}
	\caption[NNLO plot for diagram $(f)$ on a larger scale]{\textbf{NNLO plot for diagram $\boldsymbol{(f)}$ on a larger scale} including the $S$-series, the $T$-series, the extrapolation of the $S$-series, the Pad\'{e} approximation of the $S$-series and the interpolation of the $S$- and $T$-series. The $S$-series and the extrapolation (the $T$-series and the interpolation on the right-hand side) are plotted up to $\mathcal{O}(x^{\pm7})$, $\mathcal{O}(x^{\pm8})$ and $\mathcal{O}(x^{\pm9})$ corresponding to the three curves from top to bottom (from bottom to top), respectively. The $[4,3]$, $[4,4]$ and $[5,4]$ Pad\'{e} approximations correspond to the three curves from top to bottom.}
	\label{fig:mixeddivplot1zoom}
	\end{center}
\end{figure}

\begin{figure}[tb]
	\begin{center}
	\includegraphics[scale=0.67]{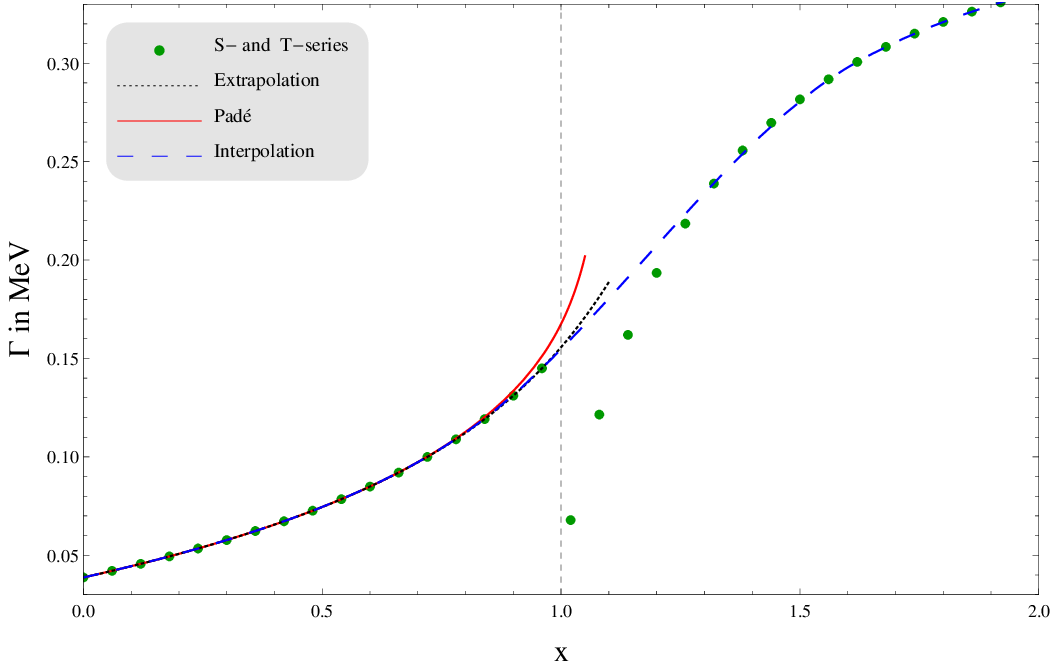}
	\caption[Combined NNLO plot for the diagrams $(g)$ and $(h)$]{\textbf{Combined NNLO plot for the diagrams $\boldsymbol{(g)}$ and $\boldsymbol{(h)}$} including the \mbox{$S$-series}, the $T$-series, the extrapolation of the $S$-series, the $[5,4]$ Pad\'{e} approximation of the $S$-series and the interpolation of the $S$- and $T$-series. Every series is plotted up to $\mathcal{O}(x^{\pm9})$.}
	\label{fig:mixeddivplot2}
	\end{center}
\end{figure}

\begin{figure}[tb]
	\begin{center}
	\includegraphics[scale=0.67]{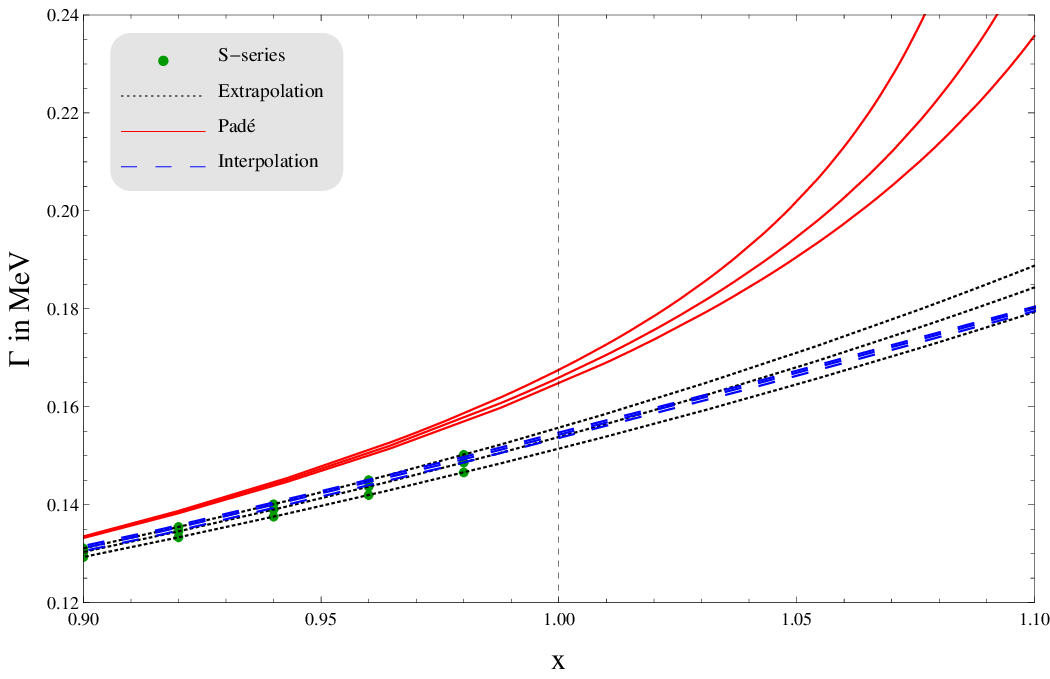}
	\caption[Combined NNLO plot for the diagrams $(g)$ and $(h)$ on a larger scale]{\textbf{Combined NNLO plot for the diagrams $\boldsymbol{(g)}$ and $\boldsymbol{(h)}$ on a larger scale} including the $S$-series, the extrapolation of the $S$-series, the Pad\'{e} approximation of the $S$-series and the interpolation of the $S$- and $T$-series. The $S$-series, the extrapolation and the interpolation are plotted up to $\mathcal{O}(x^{\pm7})$, $\mathcal{O}(x^{\pm8})$ and $\mathcal{O}(x^{\pm9})$ corresponding to the three curves from bottom to top, respectively. The $[4,3]$, $[4,4]$ and $[5,4]$ Pad\'{e} approximations correspond to the three curves from bottom to top.}
	\label{fig:mixeddivplot2zoom}
	\end{center}
\end{figure}

We have calculated the coefficients up to $\mathcal{O}(\delta^{5})$ so that a statement about the convergence can be made with the help of Table~\ref{tab:mixedtaylorconv}. From there, we can deduce that the Taylor series in the mass difference of the two bosons at NNLO converges sufficiently fast within the accuracy of the final result.\\
We use the results in Eqs.~\eqref{mixedasymp} to study the convergence of the asymptotic expansion by examining the difference of the $S$-series including powers up to $x^{j+1}$ and $x^j$ for increasing~$j$. Again, this has been done for $\delta = 0$ since the Taylor expansion in $\delta$ has proven to converge:
\begin{align}
\label{mixedasympconv}
	\Delta_j S^{(2)} &\equiv \Gamma^{(0)} \, \frac{\alpha}{\pi} \, \frac{\alpha_s}{\pi}\, \left( \sum_{n=0}^j d_n \, x^n - \sum_{n=0}^{j-1} d_n \, x^n \right) \notag \\
	&= \Gamma^{(0)} \, \frac{\alpha}{\pi} \, \frac{\alpha_s}{\pi} \, d_j \, x^j \hspace{4cm} (j = 1..9) \, .
\end{align}
The convergent diagrams can be separated from the slowly converging ones by means of Table~\ref{tab:mixedasympconv}, which shows the values of $\Delta_j S^{(2)}$ for each group~$(a)-(g)$. As at NLO, the groups~$(a)-(d)$ belong to the convergent ones whereas $(f)$ and $(g)$ converge extremely slowly. However, the coefficients $\Delta_j S^{(2)}$ for diagram $(e)$ lie in the same range as the values for $(f)$ and $(g)$ so that it will be counted amongst the slowly converging diagrams in contrast to the NLO calculation.
\begin{table}[tb]
\caption[Convergence of the asymptotic expansion in $x$ for the various NNLO contributions]{\textbf{Convergence of the asymptotic expansion in $\boldsymbol{x}$ for the various NNLO contributions} expressed through $\Delta_j S^{(2)}$ as defined in Eq.~\eqref{mixedasympconv} for each group~$(a)-(g)$ and their sum $\sum$. All quantities are indicated for $\left| V_{qq'} \right|^2 = 1$. The numerical values are given in keV and the input parameters can be found in Section~\ref{sec:num}.}
\setlength{\tabcolsep}{0.175cm}
\begin{tabularx}{1\textwidth}{cclllrrrr}
\toprule
 & \multicolumn{1}{c}{$(a)$} & \multicolumn{1}{c}{$(b)$} & \multicolumn{1}{c}{$(c)$} & \multicolumn{1}{c}{$(d)$} & \multicolumn{1}{c}{$(e)$} & \multicolumn{1}{c}{$(f)$} & \multicolumn{1}{c}{$(g)$} & \multicolumn{1}{c}{$\sum$} \\
\midrule
$\Delta_2 S^{(2)}$ & $0$ & $-8.7$ & $-3.0$ & $-1.2$ & $+2.4$ & $-29.1$ & $-3.8$ & $-43.4$ \\
$\Delta_3 S^{(2)}$ & $0$ & $-1.3$ & $-0.2$ & $-9.7 \cdot 10^{-2}$ & $+17.4$ & $-15.3$ & $-6.4$ & $-5.9$ \\
$\Delta_4 S^{(2)}$ & $0$ & $-0.2$ & $-3.8 \cdot 10^{-2}$ & $-1.5 \cdot 10^{-2}$ & $-12.1$ & $-9.4$ & $-6.7$ & $-28.6$ \\
$\Delta_5 S^{(2)}$ & $0$ & $-4.1 \cdot 10^{-2}$ & $-8.6 \cdot 10^{-3}$ & $-3.4 \cdot 10^{-3}$ & $+9.4$ & $-6.4$ & $-6.4$ & $-3.5$ \\
$\Delta_6 S^{(2)}$ & $0$ & $-8.1 \cdot 10^{-3}$ & $-2.4 \cdot 10^{-3}$ & $-9.5 \cdot 10^{-4}$ & $-6.6$ & $-4.6$ & $-6.0$ & $-17.2$ \\
$\Delta_7 S^{(2)}$ & $0$ & $-1.6 \cdot 10^{-3}$ & $-8.2 \cdot 10^{-4}$ & $-3.2 \cdot 10^{-4}$ & $+5.0$ & $-3.5$ & $-5.5$ & $-4.1$ \\
$\Delta_8 S^{(2)}$ & $0$ & $-3.3 \cdot 10^{-4}$ & $-3.2 \cdot 10^{-4}$ & $-1.2 \cdot 10^{-4}$ & $-3.7$ & $-2.8$ & $-5.1$ & $-11.7$ \\
$\Delta_9 S^{(2)}$ & $0$ & $-7.2 \cdot 10^{-5}$ & $-1.4 \cdot 10^{-4}$ & $-5.3 \cdot 10^{-5}$ & $+3.0$ & $-2.2$ & $-4.8$ & $-4.1$ \\
\midrule
$S^{(2)}$ & \multicolumn{1}{c}{$-36.6$} & \multicolumn{1}{c}{$-421.2$} & \multicolumn{1}{c}{$10.9$} & \multicolumn{1}{c}{$4.3$} & \multicolumn{1}{c}{$18.6$} & \multicolumn{1}{c}{$-188.3$} & \multicolumn{1}{c}{$-44.8$} & \multicolumn{1}{c}{$-657.0$} \\
\bottomrule
\label{tab:mixedasympconv}
\end{tabularx}
\end{table}
Accordingly, the convergent diagrams' contribution to the mixed QCD/electroweak corrections is given by the sum of the $S$-series of the groups~$(a)-(d)$: 
\begin{equation}
	\label{mixedwidthconv}
	\Gamma^{(2)}_\mathrm{conv} = \sum_{i=a}^d \Gamma^{(1)}_i = \lim_{x \to 1} \, \sum_{i=a}^d S^{(2)}_i \, .
\end{equation}
The contribution of the slowly converging diagrams to the mixed QCD/electroweak corrections stems from three parts corresponding to the groups~$(e)$, $(f)$ and $(g)$:
\begin{equation}
	\label{mixedwidthdiv}
	\Gamma^{(2)}_\mathrm{slow} = \Gamma^{(2)}_e + \Gamma^{(2)}_f + \Gamma^{(2)}_g \, .
\end{equation}
The determination of $\Gamma^{(2)}_f$ and $\Gamma^{(2)}_g$ is based on the approximation methods described in Section \ref{sec:ana2}, whose prescriptions have been developed with the help of the NLO on-shell results. These prescriptions have been applied to the NNLO calculation, which results in Figs.~\ref{fig:mixeddivplot1}~-~\ref{fig:mixeddivplot2zoom}. Besides the $S$-series, we have computed the Pad\'{e} approximant of the $S$-series as well as the $T$-series in order to obtain them:
\begin{align}
	P^{(2)}_f [5,4] &= \frac{-1.37 \cdot 10^{-4} \, x^5 +9.11 \cdot 10^{-3} \, x^4 -0.056 \, x^3 +0.090 \, x^2 -0.014 \, x -0.033}{0.026 \, x^4 -0.349 \, x^3 +1.371 \, x^2 -2.025 \, x +1} \, ,\\
	t^{(2)}_f &= \frac{77}{24} - \frac{20}{3} \, \zeta(3) + \frac{7}{12} \, \frac{1}{\epsilon} + \frac{7}{4} \, \mathrm{ln} \, \frac{\mu^2}{M_W^2} - \frac{17}{4} \, \mathrm{ln} \, x  \notag \\
& \, \quad + \left(\frac{325}{48} - 7 \, \zeta(3) + \frac{7}{8} \, \frac{1}{\epsilon} + \frac{21}{8} \, \mathrm{ln} \, \frac{\mu^2}{M_W^2} - 5 \, \mathrm{ln} \, x \right) \frac{1}{x}  \notag \\
& \, \quad + \left(-\frac{61}{432} + \frac{5}{3} \, \zeta(3) - \frac{5}{24} \, \frac{1}{\epsilon} - \frac{5}{8} \, \mathrm{ln} \, \frac{\mu^2}{M_W^2} + \frac{1}{36} \, \mathrm{ln} \, x \right) \frac{1}{x^2} + \left(\frac{151}{864} - \frac{25}{72} \, \mathrm{ln} \, x \right) \frac{1}{x^3} \notag \\
\displaybreak
& \, \quad + \left(\frac{3817}{21600} - \frac{23}{90} \, \mathrm{ln} \, x \right) \frac{1}{x^4} + \left(\frac{2773}{21600} - \frac{73}{360} \, \mathrm{ln} \, x \right) \frac{1}{x^5} \label{mixedtf} \\
& \, \quad + \left(\frac{99823}{1058400} - \frac{53}{315} \, \mathrm{ln} \, x \right) \frac{1}{x^6} + \left(\frac{15149}{211680} - \frac{145}{1008} \, \mathrm{ln} \, x \right) \frac{1}{x^7} \notag \\
& \, \quad + \left(\frac{42647}{762048} - \frac{95}{756} \, \mathrm{ln} \, x \right) \frac{1}{x^8} + \left(\frac{15271}{340200} - \frac{241}{2160} \, \mathrm{ln} \, x \right) \frac{1}{x^9} + \mathcal{O}\left(\frac{1}{x^{10}} \right) \, , \tag{\ref{mixedtf}} \\
	t^{(2)}_g &= \frac{159}{16} - \frac{11}{3} \, \zeta(3) + \frac{11}{24} \, \frac{1}{\epsilon} + \frac{11}{8} \, \mathrm{ln} \, \frac{\mu^2}{M_W^2} - \frac{11}{8} \, \mathrm{ln} \, x \notag \\
& \, \quad + \left(-\frac{217}{16} + 11 \, \zeta(3) - \frac{11}{8} \, \frac{1}{\epsilon} - \frac{33}{8} \, \mathrm{ln} \, \frac{\mu^2}{M_W^2} + \frac{33}{8} \, \mathrm{ln} \, x \right) \frac{1}{x} \notag \\
& \, \quad + \left(-\frac{17}{24} - \frac{10}{3} \, \zeta(3) + \frac{5}{12} \, \frac{1}{\epsilon} + \frac{5}{4} \, \mathrm{ln} \, \frac{\mu^2}{M_W^2} - \frac{5}{4} \, \mathrm{ln} \, x \right) \frac{1}{x^2} -\frac{5}{8} \frac{1}{x^3} - \frac{19}{30} \, \frac{1}{x^4} \notag \\
& \, \quad - \frac{37}{72} \, \frac{1}{x^5} - \frac{59}{140} \, \frac{1}{x^6} - \frac{17}{48} \, \frac{1}{x^7} - \frac{115}{378} \, \frac{1}{x^8} - \frac{149}{560} \, \frac{1}{x^9} + \mathcal{O}\left(\frac{1}{x^{10}} \right) \, .
\end{align}

As for the $S$- and $s$-series, we have introduced the abbreviated form of the inverse asymptotic expansion:
\begin{equation}
	T^{(2)} = \Gamma^{(0)} \, \frac{\alpha}{\pi} \, \frac{\alpha_s}{\pi} \, \sum_{n=0}^{\infty} \frac{\tilde{d}_n}{x^n} \equiv \Gamma^{(0)} \, \frac{\alpha}{\pi} \, \frac{\alpha_s}{\pi} \, t^{(2)} \, .
\end{equation}
According to Section \ref{sec:ana2}, we have to combine group~$(g)$ with the photonic contribution to the field renormalization constant $\delta Z_W$ of the $W$ boson (Fig.~\ref{fig:deltaZW}). For this purpose, the \mbox{$S$- and $T$-series} of group~$(h)$ have to be evaluated at NNLO. They immediately follow from the NLO expressions (Eqs.~\eqref{dZWgammaS} and \eqref{dZWgammaT}) by multiplying the coefficients of the asymptotic expansion by the NLO QCD corrected width up to $\mathcal{O}(\epsilon)$ (Eq.~\eqref{qcdeps}) instead of the Born width up to $\mathcal{O}(\epsilon)$ (Eq.~\eqref{borneps}). Similarly, the on-shell result of $(h)$ at NNLO can be derived from Eq.~\eqref{dZWgamma}:
\begin{align}
	\Gamma^{(2)}_h &= \frac{\Gamma^{(1)}_{\mathrm{QCD},\epsilon}}{\Gamma^{(0)}_\epsilon} \, \Gamma^{(1)}_h \label{dZWgamma2} \, ,\\
	S^{(2)}_h &= \frac{\Gamma^{(1)}_{\mathrm{QCD},\epsilon}}{\Gamma^{(0)}_\epsilon} \, S^{(1)}_h \, ,\\
	T^{(2)}_h &= \frac{\Gamma^{(1)}_{\mathrm{QCD},\epsilon}}{\Gamma^{(0)}_\epsilon} \, T^{(1)}_h \label{dZWgammaT2} \, .
\end{align}
The numerical result of the combination $(g)+(h)$ is then determined by the value of the Pad\'{e} approximant
\begin{equation}
	P^{(2)}_{g+h} [5,4] = \frac{-2.38 \cdot 10^{-4} \, x^5 -2.57 \cdot 10^{-3} \, x^4 +0.024 \, x^3 -0.032 \, x^2 -0.024 \, x +0.039}{0.027 \, x^4 -0.357 \, x^3 +1.385 \, x^2 -2.033 \, x +1} \, .
\end{equation}
As stated in Section \ref{sec:ana2}, we will use the Pad\'{e} approximated on-shell results for the numerical evaluation, i.e. the value of their curves in Figs.~\ref{fig:mixeddivplot1} and \ref{fig:mixeddivplot2} for $x=1$. The result for group~$(g)$ can then be obtained by substracting the on-shell result for $(h)$ from the combined value $(g)+(h)$:
\begin{equation}
	\Gamma^{(2)}_g = \Gamma^{(2)}_{g+h} - \Gamma^{(2)}_h \, .
\end{equation}
\clearpage
\clearpage
What remains to be found is a solution for group~$(e)$, for which we will use another method proposed by Refs. \cite{Czarnecki:1996, Seidensticker:1998}: From Eqs.~\eqref{mixedasymp}, we read off that the results of the groups~$(d)$ and $(e)$ are proportional to $g_q^2 + g_{q'}^2$ and $g_q \, g_{q'}$, respectively. The relationships
\begin{equation}
	\underbrace{\vphantom{\frac{c_w}{s_w}}g_q^2}_{(d)} = \underbrace{2 \, \frac{c_w}{s_w} \, I_q^3 \, g_q}_{(d1)} + \underbrace{\vphantom{\frac{c_w}{s_w}}g_q \, g_{q'}}_{(d2)} \qquad \text{and} \qquad \underbrace{\vphantom{\frac{c_w}{s_w}}g_{q'}^2}_{(d)} = \underbrace{2 \, \frac{c_w}{s_w} \, I_{q'}^3 \, g_{q'}}_{(d1)} + \underbrace{\vphantom{\frac{c_w}{s_w}}g_q \, g_{q'}}_{(d2)}
\end{equation}
follow from the invariance of the weak hypercharge $Y = \nicefrac{1}{2} \, (Q_q - I_q^3)$ with respect to the isospin. They can be used to split the prefactor of $s^{(2)}_d$ into two parts, from which two new series arise. The first one is proportional to $c_w/s_w \, (g_q+g_{q'})$; we call it $s^{(2)}_{d1}$ and find that it keeps the satisfactory convergence behavior of $s^{(2)}_d$. The prefactor of the second part $s^{(2)}_{d2}$ equals that of $s^{(2)}_e$ so that we can consider their sum:
\begin{align}
	s^{(2)}_{d2+e} &= g_q \, g_{q'} \left[ \left(\frac{397}{243} - \frac{4}{3} \, \zeta(3) - \frac{10}{81} \, \mathrm{ln} \, x - \frac{11}{27} \, \mathrm{ln}^2 \, x\right) x \right. \notag \\
& \qquad \qquad + \left(-\frac{1823}{3888} + \frac{1}{3} \, \zeta(3) + \frac{553}{648} \, \mathrm{ln} \, x - \frac{43}{216} \, \mathrm{ln}^2 \, x \right) x^2 \notag \\
& \qquad \qquad + \left(-\frac{421343}{2430000} - \frac{2}{15} \, \zeta(3) - \frac{1439}{10125} \, \mathrm{ln} \, x - \frac{1}{1350} \, \mathrm{ln}^2 \, x \right) x^3 \notag \\
& \qquad \qquad + \left(\frac{364763}{2430000} + \frac{1}{15} \, \zeta(3) + \frac{2779}{27000} \, \mathrm{ln} \, x - \frac{331}{5400} \, \mathrm{ln}^2 \, x \right) x^4 \notag \\
& \qquad \qquad + \left(-\frac{32249507}{243101250} - \frac{4}{105} \, \zeta(3) - \frac{44897}{3472875} \, \mathrm{ln} \, x - \frac{2}{33075} \, \mathrm{ln}^2 \, x \right) x^5 \notag \\
& \qquad \qquad  + \left(\frac{1799816677}{18670176000} + \frac{1}{42} \, \zeta(3) + \frac{50657}{2963520} \, \mathrm{ln} \, x - \frac{6011}{211680} \, \mathrm{ln}^2 \, x \right) x^6 \notag \\
& \qquad \qquad + \left(-\frac{115885580383}{1512284256000} - \frac{1}{63} \, \zeta(3) + \frac{176657}{30005640} \, \mathrm{ln} \, x - \frac{223}{238140} \, \mathrm{ln}^2 \, x \right) x^7 \notag \\
& \qquad \qquad + \left(\frac{62579290843}{1080203040000} + \frac{1}{90} \, \zeta(3) + \frac{16201}{122472000} \, \mathrm{ln} \, x - \frac{21769}{1360800} \, \mathrm{ln}^2 \, x \right) x^8 \notag \\
& \qquad \qquad \left. + \left(-\frac{46158233533877}{988453294290000} - \frac{4}{495} \, \zeta(3) + \frac{20762534}{2547034875} \, \mathrm{ln} \, x - \frac{12007}{10291050} \, \mathrm{ln}^2 \, x \right) x^9 \right] \notag \\
& \qquad \qquad + \mathcal{O}\left(x^{10} \right) \, . \label{d2e}
\end{align}
Hence, this sum is finite and its lowest-order coefficient $d_0$ vanishes. Let us repeat the exact same procedure for the inverse asymptotic expansion:
\begin{align}
	t^{(2) \, \mathrm{CK}}_{d2+e} &= g_q \, g_{q'} \left[ -\frac{1}{4} + \frac{1}{4} \, \frac{1}{x^2} \right. \label{excep} \\
& \qquad \qquad \left. + \left(\frac{397}{243} - \frac{4}{3} \, \zeta(3) + \frac{10}{81} \, \mathrm{ln} \, x - \frac{11}{27} \, \mathrm{ln}^2 \, x \right) \frac{1}{x^3} \right] + \mathcal{O}\left(\frac{1}{x^4}\right) \notag \, .
\end{align}
We have added the superscript `CK' to this series for the following reason: Its para\-me\-terization follows the one by A. Czarnecki and J. H. K\"uhn in Refs. \cite{Czarnecki:1996, Seidensticker:1998} and differs from the para\-me\-terization we have used so far (without the superscript). The coefficients $\tilde{c}_n$ and $\tilde{c}_n^{\, \mathrm{CK}}$ of both series as defined in Eq.~\eqref{ewasympnotationinv} are connected as follows:
\begin{equation}
	\tilde{c}_n =
\left\{
  \begin{array}{ll}
    \tilde{c}_{n+1}^{\, \mathrm{CK}} & \qquad (n=1..10) \, , \\
	& \\
    \tilde{c}_{0}^{\, \mathrm{CK}} + \tilde{c}_{1}^{\, \mathrm{CK}} & \qquad (n=0) \, .
  \end{array}
\right.
\label{para}
\end{equation}
The higher-order coefficients of the inverse asymptotic series in Eq.~\eqref{excep} have been omitted due to an exceptional feature: Starting from $n=3$, the coefficients of $\mathcal{O}(1/x^n)$ in $t^{(2) \, \mathrm{CK}}_{d2+e}$ and the coefficients of $\mathcal{O}(x^{n-2})$ in $s^{(2)}_{d2+e}$ agree. We can benefit from that in order to find an approximated value for the on-shell result of diagram $(e)$ as follows: The behavior of the coefficients ensures that both the $s^{(2)}_{d2+e}$- and the $t^{(2) \, \mathrm{CK}}_{d2+e}$-series yield the same value at the threshold $x=1$. Consequently, $t^{(2) \, \mathrm{CK}}_{d2+e}$ can be used instead of $s^{(2)}_{d2+e}$ to determine the on-shell result. Since the coefficient of $\mathcal{O}(1/x)$ within the $t^{(2) \, \mathrm{CK}}_{d2+e}$-series vanishes, it hardly depends on $x$ over a wide range. Therefore, the on-shell result for $x=1$ can be approximated by its value for $x \rightarrow \infty$:
\begin{equation}
	\Gamma^{(2)}_{d2+e} \approx \lim_{x \to \infty} T^{(2) \, \mathrm{CK}}_{d2+e} = -\Gamma^{(0)} \, \frac{\alpha}{\pi} \, \frac{\alpha_s}{\pi} \frac{1}{4} \, g_q \, g_{q'} \, .
\end{equation}
The sum of the contributions of the groups~$(d)$ and $(e)$ is then given by adding $\Gamma^{(2)}_{d1}$, which is associated with $s^{(2)}_{d1}$, to $\Gamma^{(2)}_{d2+e}$. Subsequently, $\Gamma^{(2)}_d$ has to be substracted in order to obtain $\Gamma^{(2)}_e$:
\begin{equation}
	\Gamma^{(2)}_e = \Gamma^{(2)}_{d2+e} + \Gamma^{(2)}_{d1} - \Gamma^{(2)}_{d} \, .
\end{equation}
As a cross check, we have applied the three approximation methods to diagram $(e)$ as well and have found agreement within uncertainties.\\
Finally, the contribution of the renormalization constants to the NNLO decay width can be separated into two parts:
\begin{equation}
	\Gamma^{(2)}_\mathrm{ren} = \underbrace{\vphantom{\Gamma^{(1)}_{\mathrm{QCD},\epsilon}}\Gamma^{(0)}_\epsilon \, \delta Z_q}_{\text{quarks}} + \underbrace{\Gamma^{(1)}_{\mathrm{QCD},\epsilon} \, \left( \delta Z_l + \delta Z_b \right)}_{\text{leptons + bosons}} \, .
\label{renren}
\end{equation}
$\delta Z_q$ contains the two-loop diagrams obtained by adding one gluon line in every possible way to the diagrams with quarks inside the loop within $\delta Z$ in Eq. \eqref{Denner3}. We find that the pole part of $\delta Z_q$ vanishes and that its finite part is given by $\alpha_s/\pi$ times the contributions of the quarks to the one-loop self-energies $\delta Z$. The second term in Eq. \eqref{renren} includes the QCD corrected width expanded up to $\mathcal{O}(\epsilon)$ times the one-loop self-energies with leptons and bosons inside the loop. Altogether, only one-loop self-energies remain, which have already been computed in order to renormalize the NLO results, thus Eq. \eqref{renren} can be rewritten as
\begin{equation}
	\Gamma^{(2)}_\mathrm{ren} = \frac{\Gamma^{(1)}_{\mathrm{QCD},\epsilon}}{\Gamma^{(0)}_\mathrm{\epsilon}} \, \Gamma^{(1)}_\mathrm{ren} \, .
\label{RenNNLO}
\end{equation}

\section{Numerical Results}
\label{sec:num}

For our numerical analysis, we use the following input parameters \cite{Beringer:2012, CMS:2012, ATLAS:2012}:
\begin{alignat}{4}
	\alpha_s^{(5)} (M_W) &= 0.120597 \, , &\quad \alpha &= 1/137.035999074 \, , &\quad G_F &= 1.1663787 \cdot 10^{-5} \, \mathrm{GeV}^{-2} \, , \notag \\
	M_W &= 80.385 \, \mathrm{GeV} \, , &\quad M_Z &= 91.1876 \, \mathrm{GeV} \, , \notag \\
	M_H &= 126 \, \mathrm{GeV} \, ,  &\quad m_t &= 173.5 \, \mathrm{GeV} \, , \notag \\
	\lvert V_{ud} \rvert &= 0.97425	\, , &\quad \lvert V_{us} \rvert &= 0.2252 \, , &\quad \lvert V_{ub} \rvert &= 4.15 \cdot 10^{-3} \, , \notag \\
	\lvert V_{cd} \rvert &= 0.23 \, ,	&\quad \lvert V_{cs} \rvert &= 1.006 &\quad \lvert V_{cb} \rvert &= 40.9 \cdot 10^{-3} \, . \label{inputparameters}
\end{alignat}
The Higgs boson mass $M_H$ and the pole mass $m_t$ of the top quark occur in the calculation of the renormalization constant in Eq. \eqref{Denner3}.
\clearpage
\noindent The value of $\alpha_s^{(5)} (M_W)$ stems from the one-loop relation
\begin{equation}
	\alpha_s^{(n_f)} (\mu) = \frac{\alpha_s^{(n_f)} (M_Z)}{1+\alpha_s^{(n_f)} (M_Z) \, \beta_0 \, \mathrm{ln}(\mu^2/M_Z^2)/\pi} \, ,
\label{running}
\end{equation}
which is evaluated at the renormalization scale $\mu = M_W$ using $\beta_0 = 11/4-n_f/6$, $\alpha_s^{(5)} (M_Z) = 0.1184$ and $n_f = 5$ active quark flavors.\\
Substituting these input parameters into the analytical results yields the values in Table~\ref{tab:NumVarious}.\footnote{The reader can inject his preferred CKM matrix elements $\tilde{V}_{qq'}$ by multiplying the values of Table~\ref{tab:NumVarious} by $\left| \tilde{V}_{qq'} \right|^2/\left| V_{qq'} \right|^2$, where $\left| V_{qq'} \right|$ is indicated in Eq. \eqref{inputparameters}. This is due to the fact that the CKM matrix is taken to be equal to the unit matrix when we calculate higher-order corrections.} It also contains the higher-order QCD corrections, which have been computed with the help of Ref. \cite{Baikov:2008}.\footnote{Note that the evaluation of the higher-order QCD corrections actually requires extending Eq. \eqref{running} to more than one loop. We have passed on that because the one-loop relation \eqref{running} leads to well-approximated results for the higher-order QCD corrections. This is due to the short distance of the running coupling originating from the small difference of the $W$ and the $Z$ boson mass. Beyond that, this work focuses on the mixed QCD/electroweak corrections, for which Eq. \eqref{running} is sufficient.} We could not only reproduce the well-known numerical results of the Born decay width $\Gamma^{(0)}$ and the NLO QCD corrections $\Gamma^{(1)}_\mathrm{QCD}$, but also the values of $\Gamma^{(1)}_\mathrm{EW}$. They have been compared to those of Ref.~\cite{Kniehl:2000} by replacing our input para\-meters with the ones used therein. In doing so, we have found agreement to an impressive accuracy of $0.01$\%. It should be stressed that the results of Ref.~\cite{Kniehl:2000} emerge from a completely different kind of calculation. The negligible deviation is assumed to be caused by the diagonal CKM matrix within the calculation of higher-order corrections and renormalization constants, by the use of massless quarks as well as by the application of asymptotic expansions and Taylor expansions.\\
The so far unknown result $\Gamma^{(2)}_\mathrm{mixed}$ of the NNLO mixed QCD/electroweak decay width is shown in Table~\ref{tab:NumVarious} as well. We observe that it is of the same order of magnitude as $\Gamma^{(3)}_\mathrm{QCD}$ and $\Gamma^{(4)}_\mathrm{QCD}$ and lies in between. In addition, we read off that the NNLO mixed QCD/electroweak contribution equals
\begin{equation}
	\frac{\Gamma^{(2)}_\mathrm{mixed}}{\Gamma^{(1)}_\mathrm{EW}} = 15.18\%
\label{ratio1}
\end{equation}
of the NLO electroweak corrections. Consequently, the actual result is four times as large as the naively expected one, which emerges from a multiplication of the NLO electroweak contribution by the QCD correction factor $\alpha_s/\pi$:
\begin{equation}
	\Gamma^{(2)}_\mathrm{mixed} \approx 4 \cdot \Gamma^{(1)}_\mathrm{EW} \cdot \frac{\alpha_s}{\pi} \, .
\label{four}
\end{equation}

\begin{table}[b]
\caption[Contributions to the numerical result of the hadronic decay width]{\textbf{Contributions to the numerical result of the hadronic decay width} according to Eq.~\eqref{OverallWidth} including the Born decay width $\Gamma^{(0)}$, the $\mathcal{O}(\alpha_s^i)$ QCD corrections $\Gamma^{(i)}_\mathrm{QCD}$, the $\mathcal{O}(\alpha)$ electroweak corrections $\Gamma^{(1)}_\mathrm{EW}$ and the $\mathcal{O}(\alpha \, \alpha_s)$ mixed QCD/electroweak corrections $\Gamma^{(2)}_\mathrm{mixed}$. All values are given in MeV.}
\setlength{\tabcolsep}{0.225cm}
\begin{tabularx}{\textwidth}{lrrccccc}
\toprule
Partial width & \multicolumn{1}{c}{$\Gamma^{(0)}$} & \multicolumn{1}{c}{$\Gamma^{(1)}_\mathrm{QCD}$} & \multicolumn{1}{c}{$\Gamma^{(2)}_\mathrm{QCD}$} & \multicolumn{1}{c}{$\Gamma^{(3)}_\mathrm{QCD}$} & \multicolumn{1}{c}{$\Gamma^{(4)}_\mathrm{QCD}$} & \multicolumn{1}{c}{$\Gamma^{(1)}_\mathrm{EW}$} & \multicolumn{1}{c}{$\Gamma^{(2)}_\mathrm{mixed}$} \\
\midrule
$\Gamma \, (W \rightarrow ud)$ & $647.158$ & $24.843$ & $1.344$ & $-0.467$ & $-0.112$ & $-2.357$ & $-0.358$ \\
$\Gamma \, (W \rightarrow us) \times 10$ & $345.785$ & $13.274$ & $0.718$ & $-0.250$ & $-0.060$ & $-1.259$ & $-0.191$ \\
$\Gamma \, (W \rightarrow ub) \times 10^4$ & $117.426$ & $4.508$ & $0.244$ & $-0.085$ & $-0.020$ & $-0.428$ & $-0.065$ \\
$\Gamma \, (W \rightarrow cd) \times 10$ & $360.683$ & $13.846$ & $0.749$ & $-0.261$ & $-0.063$ & $-1.314$ & $-0.199$ \\
$\Gamma \, (W \rightarrow cs)$ & $690.026$ & $26.488$ & $1.433$ & $-0.468$ & $-0.120$ & $-2.513$ & $-0.382$ \\
$\Gamma \, (W \rightarrow cb) \times 10^2$ & $114.056$ & $4.378$ & $0.237$ & $-0.082$ & $-0.020$ & $-0.415$ & $-0.063$ \\
\midrule
$\Gamma \, (W \rightarrow \mathrm{hadrons})$ & $1408.980$ & $54.087$ & $2.927$ & $-1.018$ & $-0.245$ & $-5.132$ & $-0.779$ \\
\bottomrule
\label{tab:NumVarious}
\end{tabularx}
\end{table}

\begin{figure}[tb!]
	\begin{center}
	\includegraphics[scale=0.67]{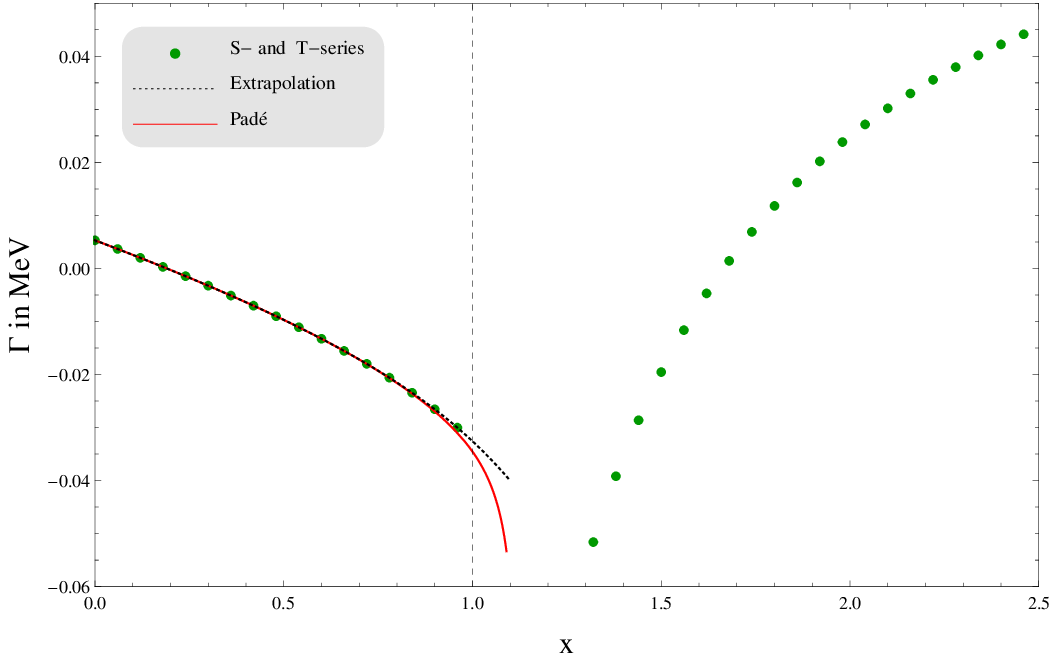}
	\caption[Combined NNLO plot for the diagrams $(f)$, $(g)$ and $(h)$]{\textbf{Combined NNLO plot for the diagrams $\boldsymbol{(f)}$, $\boldsymbol{(g)}$ and $\boldsymbol{(h)}$} including the $S$-series, the $T$-series, the extrapolation of the $S$-series, the $[5,4]$ Pad\'{e} approximation of the $S$-series and the interpolation of the $S$- and $T$-series. Every series is plotted up to $\mathcal{O}(x^{\pm9})$.}
	\label{fig:mixeddivplot12}
	\end{center}
\end{figure}

\begin{figure}[tb!]
	\begin{center}
	\includegraphics[scale=0.67]{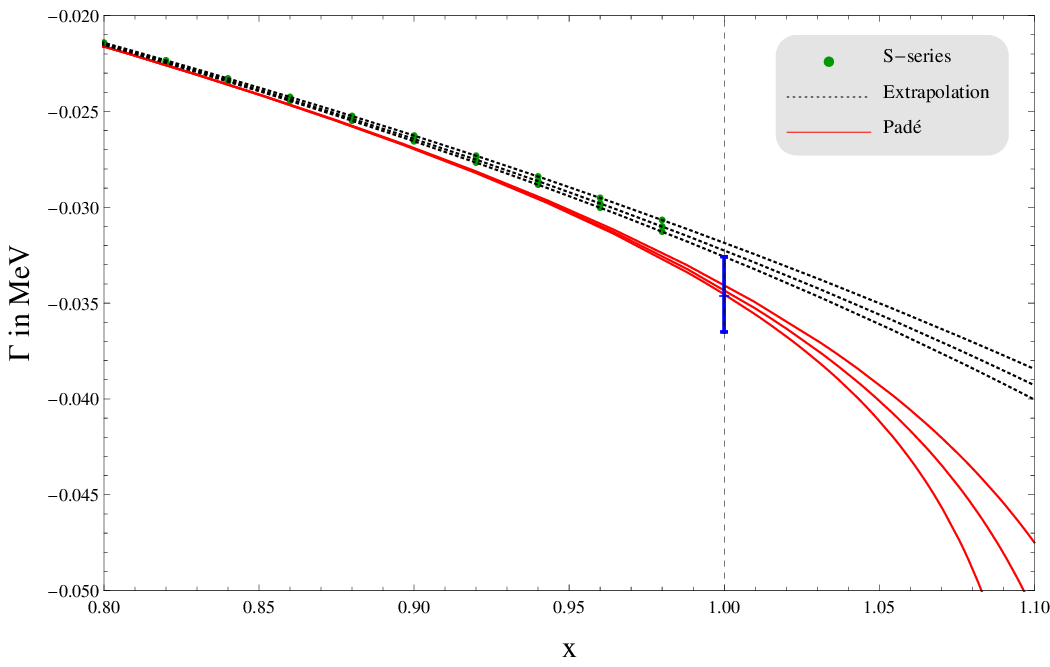}
	\caption[Combined NNLO plot for the diagrams $(f)$, $(g)$ and $(h)$ on a larger scale]{\textbf{Combined NNLO plot for the diagrams $\boldsymbol{(f)}$, $\boldsymbol{(g)}$ and $\boldsymbol{(h)}$ on a larger scale} including the $S$-series, the extrapolation of the $S$-series and the Pad\'{e} approximation of the $S$-series. The $S$-series and the extrapolation are plotted up to $\mathcal{O}(x^{\pm7})$, $\mathcal{O}(x^{\pm8})$ and $\mathcal{O}(x^{\pm9})$ corresponding to the three curves from top to bottom, respectively. The $[4,3]$, $[4,4]$ and $[5,4]$ Pad\'{e} approximations correspond to the three curves from top to bottom. The error bar serves to determine the uncertainty of the result in Section \ref{sec:num}.}
	\label{fig:mixeddivplot12zoom}
	\end{center}
\end{figure}

Finally, the so far unknown mixed QCD/electroweak corrections account for
\begin{equation}
	\frac{\Gamma^{(2)}_\mathrm{mixed}}{\Gamma_\mathrm{had}} = -0.0534\%
\label{ratio2}
\end{equation}
of the overall hadronic decay width as defined in Eq.~\eqref{OverallWidth}.\\
In a final step, we would like to examine the uncertainty associated with the computation of the mixed QCD/electroweak corrections. In case of group~$(e)$, Refs. \cite{Czarnecki:1996, Seidensticker:1998} propose to estimate the error of this method by evaluating the \mbox{$t^{(2) \, \mathrm{CK}}_{d2+e}$-series} in Eq.~\eqref{excep} near the threshold, e.g. for $x = M_Z^2/M_W^2 \approx 1.29$. This leads to a deviation of about
\begin{equation}
	\frac{\Delta \Gamma^{(2)}_e}{\Gamma^{(2)}_e} \approx 10\% \, .
\label{deviatione}
\end{equation}
Concerning the groups~$(f)$ and $(g)$, we choose to estimate the uncertainties of the results with the help of their combined NNLO plots (Figs.~\ref{fig:mixeddivplot12} and \ref{fig:mixeddivplot12zoom}).
The Pad\'{e} approximant of this combination reads
\begin{equation}
	P^{(2)}_{f+g+h} [5,4] = \frac{-3.07 \cdot 10^{-4} \, x^5 +6.03 \cdot 10^{-3} \, x^4 -0.030 \, x^3 +0.056 \, x^2 -0.038 \, x +5.31 \cdot 10^{-3}}{0.021 \, x^4 -0.320 \, x^3 +1.313 \, x^2 -1.990 \, x +1} \, ,
\end{equation}
leading to
\begin{equation}
	\Gamma^{(2)}_{f+g} \, (W \rightarrow \mathrm{hadrons}) \equiv \Gamma^{(2)}_f + \Gamma^{(2)}_g = -26.6 \, \mathrm{keV} \, .
\end{equation}
If we had used extrapolation instead, we would have obtained
\begin{equation}
	\Gamma^{(2)}_{f+g,\text{Ext}} \, (W \rightarrow \mathrm{hadrons}) = -22.6 \, \mathrm{keV} \, .
\end{equation}
The difference of these two numbers corresponds to the gap of the curves associated with the Pad\'{e} approximation and the extrapolation at $x=1$ in Fig.~\ref{fig:mixeddivplot12}. As depicted by the error bar in Fig.~\ref{fig:mixeddivplot12zoom}, we suppose that the uncertainty of the approximation methods equals that gap, yielding
\begin{equation}
	\Delta \Gamma^{(2)}_{f+g} \, (W \rightarrow \mathrm{hadrons}) = 4.0 \, \mathrm{keV} \, .
\end{equation}
The uncertainty associated with the results of the convergent diagrams $(a)-(d)$ is due to the application of asymptotic expansions. It is much smaller than the uncertainties associated with the slowly converging groups and can be neglected. The same holds for the uncertainty associated with the Taylor expansion in the mass difference of the $W$ and the $Z$~boson: In order to estimate the error of the Taylor expansion at NNLO, we omit terms of $\mathcal{O}(\delta^5)$ within $\Gamma^{(2)}_\mathrm{mixed}$. According to Table~\ref{tab:mixedtaylorconv}, this leads to an uncertainty of only $1.0 \, \mathrm{keV}$. Thus, the error is such that it will not affect the overall uncertainty of the mixed QCD/elec\-troweak corrections. Hence, the overall uncertainty is solely given by error propagation of $\Delta \Gamma^{(2)}_e$ and $\Delta \Gamma^{(2)}_{f+g}$:
\begin{equation}
	\Delta \Gamma^{(2)}_\mathrm{mixed} \, (W \rightarrow \mathrm{hadrons}) = \sqrt{\left(\Delta \Gamma^{(2)}_e \right)^2 + \left(\Delta \Gamma^{(2)}_{f+g}\right)^2} = 5.9 \, \mathrm{keV} \, .
\end{equation}
Therefore, our final results read
\begin{align}
	\Gamma^{(2)}_\mathrm{mixed} \, (W \rightarrow \mathrm{hadrons}) &= (-0.779 \pm 0.006) \, \mathrm{MeV} \, ,\\
	\Gamma_\mathrm{had} \, (W \rightarrow \mathrm{hadrons}) &= (1.458820 \pm 6 \cdot 10^{-6}) \, \mathrm{GeV} \, .
\end{align}
For Eqs.~\eqref{ratio1} and \eqref{ratio2}, this entails the following uncertainties:
\begin{align}
	\frac{\Gamma^{(2)}_\mathrm{mixed}}{\Gamma^{(1)}_\mathrm{EW}} &= (15.18 \pm 0.11)\% \, ,\\
	\frac{\Gamma^{(2)}_\mathrm{mixed}}{\Gamma_\mathrm{had}} &= (-0.0534 \pm 0.0004)\% \, .
\end{align}

\section{Conclusion}

We have applied two powerful tools, the optical theorem and asymptotic expansions, to the $W$ boson two-point function. That way, we have been able to compute the LO Born decay width and the NLO QCD corrections of $\mathcal{O}(\alpha_s)$. Subsequently, we have approached the more challenging NLO electroweak corrections of $\mathcal{O}(\alpha)$ finding that the results are in agreement with the ones in the literature. Within this calculation, we have encountered problems with respect to the convergence of the asymptotic series of certain classes of Feynman diagrams. The problems could be circumvented by computing the exact on-shell result of those diagrams. On the basis of these findings, we have applied three approximation methods in order to reproduce the exact on-shell result by making use of the asymptotic series. From that, we have acquired prescriptions for the extrapolation, the interpolation and the Pad\'{e} approximation of the asymptotic series whereof the latter has turned out to be the most suitable one. When passing from NLO to NNLO, the same types of Feynman diagrams have proven to cause difficulties concerning the convergence of the asymptotic series. Consequently, the NLO prescription has been applied to the Feynman diagrams associated with the NNLO mixed QCD/electroweak corrections of $\mathcal{O}(\alpha \, \alpha_s)$ yielding the so far unknown results. Next, we have analyzed the uncertainties associated with the approximation methods. Although they have turned out to keep within reasonable limits, one might be interested in minimizing them. As a next step, this could be done by computing the exact on-shell result of the three-loop two-point functions at NNLO as well.\\
By numerically evaluating the results, we have seen that they exceed the naively expected ones and are almost as large as the $\mathcal{O}(\alpha_s^3)$ QCD corrections. Hence, our results might play a role when dealing with high-precision measurements in the future.

\section*{Acknowledgements}
I would like to thank M. Steinhauser for the intensive supervision throughout this project and for careful reading of the manuscript. This work was supported by the BMBF through 05H12VKE.
\clearpage
\clearpage

%% ----------------------------------
%% |  Style of appendix numbering   |
%% ----------------------------------
 \renewcommand\appendix{\par 
   \setcounter{section}{0}% 
   \setcounter{subsection}{0}% 
   \setcounter{figure}{0}%
   \setcounter{equation}{0}%
   \renewcommand\thesection{\Alph{section}}% 
   \renewcommand\thefigure{\Alph{section}.\arabic{figure}} 
   \renewcommand\thetable{\Alph{section}.\arabic{table}}
   \renewcommand\theequation{\Alph{section}.\arabic{equation}}}
%% --- End of appendix numbering style ---

\clearpage

\begin{appendix}

\section{General Results: Coefficients for \texorpdfstring{$\boldsymbol{M_W \neq M_Z}$}{MWMZ}}
\label{sec:Generalresults}

As outlined in Sections~\ref{sec:ana2} and \ref{sec:ana3}, Eqs.~\eqref{ewasymp} and \eqref{mixedasymp} specify the results of the asymptotic expansions at NLO and NNLO only for the special case $M_W = M_Z$. The general results can be obtained by adding expressions of the form \eqref{ewdelta} and \eqref{mixeddelta} to every asymptotic~series with a $Z$ boson inside the loop, i.e. to the series of the groups $(b)$, $(d)$ and $(e)$ in Figs.~\ref{fig:ewdiag} and \ref{fig:mixeddiag}. The coefficients $a_{n,m}$ and $b_{n,m}$ belong to the NLO electroweak corrections of $\mathcal{O}(\alpha)$ and the NNLO mixed QCD/electroweak corrections of $\mathcal{O}(\alpha \, \alpha_s)$ and will be presented in Sections~\ref{sec:ewcoeff} and \ref{sec:mixedcoeff}, respectively. Within one group $(b)$, $(d)$ or $(e)$, these coefficients are accompanied by the same prefactor $P$, which is split off according to
\begin{equation}
	a_{n,m} = P \cdot A_{n,m} \, , \quad b_{n,m} = P \cdot B_{n,m} \, .
\end{equation}
$P$ agrees with the prefactors of the corresponding $s$-series in Eqs. \eqref{ewasymp} and \eqref{mixedasymp} and will be indicated individually.

\subsection{Next-To-Leading-Order Decay Width:\texorpdfstring{\\}{} Electroweak Corrections of Order \texorpdfstring{$\boldsymbol{\alpha}$}{alpha}}
\label{sec:ewcoeff}

\underline{Coefficients of group $(b)$: $P = c_w/s_w \, \left( g_q - g_{q'} \right)$}
\begin{align}
	A_{0,1} &= \frac{3}{4} \, , A_{0,2} = \frac{1}{4} \, , A_{0,3} = \frac{1}{8} \, , A_{0,4} = \frac{3}{40} \, , A_{0,5} = \frac{1}{20} \, , A_{0,6} = \frac{1}{28} \, , A_{0,7} = \frac{3}{112} \, , A_{0,8} = \frac{1}{48} \, , \notag \\
A_{0,9} &= \frac{1}{60} \, , A_{0,10} = \frac{3}{220} \, , \notag \\
	A_{1,1} &= -\frac{5}{36} \, , A_{1,2} = -\frac{1}{10} \, , A_{1,3} = -\frac{29}{360} \, , A_{1,4} = -\frac{43}{630} \, , A_{1,5} = -\frac{5}{84} \, , A_{1,6} = -\frac{10}{189} \, , \notag \\
A_{1,7} &= -\frac{103}{2160} \, , A_{1,8} = -\frac{43}{990} \, , A_{1,9} = -\frac{79}{1980} \, , \notag \\
	A_{2,1} &= -\frac{5}{144} \, , A_{2,2} = -\frac{9}{280} \, , A_{2,3} = -\frac{149}{5040} \, , A_{2,4} = -\frac{55}{2016} \, , A_{2,5} = -\frac{17}{672} \, , A_{2,6} = -\frac{7}{297} \, , \notag \\
A_{2,7} &= -\frac{131}{5940} \, , A_{2,8} = -\frac{237}{11440} \, , \notag \\
	A_{3,1} &= -\frac{23}{2800} \, , A_{3,2} = -\frac{1}{105} \, , A_{3,3} = -\frac{17}{1680} \, , A_{3,4} = -\frac{191}{18480} \, , A_{3,5} = -\frac{91}{880} \, , \notag \\
A_{3,6} &= -\frac{329}{32175} \, , A_{3,7} = -\frac{201}{20020} \, , \notag \\
	A_{4,1} &= -\frac{73}{37800} \, , A_{4,2} = -\frac{5}{1848} \, , A_{4,3} = -\frac{547}{166320} \, , A_{4,4} = -\frac{287}{77220} \, , A_{4,5} = -\frac{259}{64350} \, , \notag \\
A_{4,6} &= -\frac{7}{1650} \, , \notag \\
	A_{5,1} &= -\frac{53}{116424} \, , A_{5,2} = -\frac{3}{4004} \, , A_{5,3} = -\frac{445}{432432} \, , A_{5,4} = -\frac{173}{135135} \, , A_{5,5} = -\frac{3}{2002} \, , \notag \\
	A_{6,1} &= -\frac{145}{1345344} \, , A_{6,2} = -\frac{7}{34320} \, , A_{6,3} = -\frac{41}{131040} \, , A_{6,4} = -\frac{2319}{5445440} \, , \notag \\
	A_{7,1} &= -\frac{19}{741312} \, , A_{7,2} = -\frac{2}{36465} \, , A_{7,3} = -\frac{977}{10501920} \, , \notag \\
	A_{8,1} &= -\frac{241}{39382200} \, , A_{8,2} = -\frac{27}{1847560} \, , \notag \\
	A_{9,1} &= -\frac{149}{101615800} \, .
\end{align}
\underline{Coefficients of group $(d)$: $P = g_q^2 + g_{q'}^2$}
\begin{align}
	A_{0,1} &= -\frac{1}{4} \, , A_{0,2} = -\frac{1}{8} \, , A_{0,3} = -\frac{1}{12} \, , A_{0,4} = -\frac{1}{16} \, , A_{0,5} = -\frac{1}{20} \, , A_{0,6} = -\frac{1}{24} \, , A_{0,7} = -\frac{1}{28} \, , \notag \\
A_{0,8} &= -\frac{1}{32} \, , A_{0,9} = -\frac{1}{36} \, , A_{0,10} = -\frac{1}{40} \, , \notag \\
	A_{1,m} &= 0 \, , \notag \\
	A_{2,m} &= 0 \, , \notag \\
	A_{3,m} &= 0 \, , \notag \\
	A_{4,m} &= 0 \, , \notag \\
	A_{5,m} &= 0 \, , \notag \\
	A_{6,m} &= 0 \, , \notag \\
	A_{7,m} &= 0 \, , \notag \\
	A_{8,m} &= 0 \, , \notag \\
	A_{9,m} &= 0 \, .
\end{align}
\underline{Coefficients of group $(e)$: $P = g_q \, g_{q'}$}
\begin{align}
	A_{0,1} &= \frac{1}{2} \, , A_{0,2} = \frac{1}{4} \, , A_{0,3} = \frac{1}{6} \, , A_{0,4} = \frac{1}{8} \, , A_{0,5} = \frac{1}{10} \, , A_{0,6} = \frac{1}{12} \, , A_{0,7} = \frac{1}{14} \, , A_{0,8} = \frac{1}{16} \, , \notag \\
A_{0,9} &= \frac{1}{18} \, , A_{0,10} = \frac{1}{20} \, , \notag \\
	A_{1,1} &= \frac{5}{18} - \frac{1}{3} \, \mathrm{ln} \, x \, , A_{1,2} = \frac{1}{9} - \frac{1}{3} \, \mathrm{ln} \, x \, , A_{1,3} = -\frac{1}{3} \, \mathrm{ln} \, x \, , A_{1,4} = -\frac{1}{12} -\frac{1}{3} \, \mathrm{ln} \, x \, , \notag \\
A_{1,5} &= -\frac{3}{20} -\frac{1}{3} \, \mathrm{ln} \, x \, , A_{1,6} = -\frac{37}{180} -\frac{1}{3} \, \mathrm{ln} \, x \, , A_{1,7} = -\frac{319}{1260} -\frac{1}{3} \, \mathrm{ln} \, x \, , \notag \\
A_{1,8} &= -\frac{743}{2520} -\frac{1}{3} \, \mathrm{ln} \, x \, , A_{1,9} = -\frac{2509}{7560} -\frac{1}{3} \, \mathrm{ln} \, x \, , \notag \\
	A_{2,1} &= -\frac{7}{72} +\frac{1}{6} \, \mathrm{ln} \, x \, , A_{2,2} = -\frac{1}{16} +\frac{1}{4} \, \mathrm{ln} \, x \, , A_{2,3} = \frac{1}{3} \, \mathrm{ln} \, x \, , A_{2,4} = \frac{1}{12} +\frac{5}{12} \, \mathrm{ln} \, x \, , \notag \\
A_{2,5} &= \frac{11}{60} +\frac{1}{2} \, \mathrm{ln} \, x \, , A_{2,6} = \frac{107}{360} +\frac{7}{12} \, \mathrm{ln} \, x \, , A_{2,7} = \frac{533}{1260} +\frac{2}{3} \, \mathrm{ln} \, x \, , A_{2,8} = \frac{1879}{3360} +\frac{3}{4} \, \mathrm{ln} \, x \, , \notag \\
	A_{3,1} &= \frac{9}{200} -\frac{1}{10} \, \mathrm{ln} \, x \, , A_{3,2} = \frac{1}{25} -\frac{1}{5} \, \mathrm{ln} \, x \, , A_{3,3} = -\frac{1}{3} \, \mathrm{ln} \, x \, , A_{3,4} = -\frac{1}{12} -\frac{1}{2} \, \mathrm{ln} \, x \, , \notag \\
A_{3,5} &= -\frac{13}{60} -\frac{7}{10} \, \mathrm{ln} \, x \, , A_{3,6} = -\frac{73}{180} -\frac{14}{15} \, \mathrm{ln} \, x \, , A_{3,7} = -\frac{55}{84} -\frac{6}{5} \, \mathrm{ln} \, x \, , \notag \\
	A_{4,1} &= -\frac{11}{450} +\frac{1}{15} \, \mathrm{ln} \, x \, , A_{4,2} = -\frac{1}{36} +\frac{1}{6} \, \mathrm{ln} \, x \, , A_{4,3} = \frac{1}{3} \, \mathrm{ln} \, x \, , A_{4,4} = \frac{1}{12} +\frac{7}{12} \, \mathrm{ln} \, x \, , \notag \\
A_{4,5} &= \frac{1}{4} +\frac{14}{15} \, \mathrm{ln} \, x \, , A_{4,6} = \frac{191}{360} +\frac{7}{5} \, \mathrm{ln} \, x \, , \notag \\
	A_{5,1} &= \frac{13}{882} -\frac{1}{21} \, \mathrm{ln} \, x \, , A_{5,2} = \frac{1}{49} -\frac{1}{7} \, \mathrm{ln} \, x \, , A_{5,3} = -\frac{1}{3} \, \mathrm{ln} \, x \, , A_{5,4} = -\frac{1}{12} -\frac{2}{3} \, \mathrm{ln} \, x \, , \notag \\
A_{5,5} &= -\frac{17}{60} -\frac{6}{5} \, \mathrm{ln} \, x \, , \notag \\
	A_{6,1} &= -\frac{15}{1568} +\frac{1}{28} \, \mathrm{ln} \, x \, , A_{6,2} = -\frac{1}{64} +\frac{1}{8} \, \mathrm{ln} \, x \, , A_{6,3} = \frac{1}{3} \, \mathrm{ln} \, x \, , A_{6,4} = \frac{1}{12} +\frac{3}{4} \, \mathrm{ln} \, x \, , \notag \\
\displaybreak
	A_{7,1} &= \frac{17}{2592} -\frac{1}{36} \, \mathrm{ln} \, x \, , A_{7,2} = \frac{1}{81} -\frac{1}{9} \, \mathrm{ln} \, x \, , A_{7,3} = -\frac{1}{3} \, \mathrm{ln} \, x \, , \label{coeffewe} \\
	A_{8,1} &= -\frac{19}{4050} +\frac{1}{45} \, \mathrm{ln} \, x \, , A_{8,2} = -\frac{1}{100} +\frac{1}{10} \, \mathrm{ln} \, x \, , \notag \\
	A_{9,1} &= \frac{21}{6050} -\frac{1}{55} \, \mathrm{ln} \, x \, . \tag{\ref{coeffewe}}
\end{align}

\subsection{Next-To-Next-To-Leading-Order Decay Width:\texorpdfstring{\\}{} Mixed QCD/Electroweak Corrections of Order \texorpdfstring{$\boldsymbol{\alpha \, \alpha_s}$}{alpha alphas}}
\label{sec:mixedcoeff}

\underline{Coefficients of group $(b)$: $P = c_w/s_w \, \left( g_q - g_{q'} \right)$}
\begin{align}
	B_{0,1} &= \frac{1}{2} \, , B_{0,2} = \frac{1}{6} \, , B_{0,3} = \frac{1}{12} \, , B_{0,4} = \frac{1}{20} \, , B_{0,5} = \frac{1}{30} \, , \notag \\
	B_{1,1} &= -\frac{7}{36} \, , B_{1,2} = -\frac{73}{540} \, , B_{1,3} = -\frac{19}{180} \, , B_{1,4} = -\frac{11}{126} \, , B_{1,5} = -\frac{113}{1512} \, , \notag \\
	B_{2,1} &= -\frac{89}{2160} \, , B_{2,2} = -\frac{127}{3360} \, , B_{2,3} = -\frac{13}{378} \, , B_{2,4} = -\frac{95}{3024} \, , B_{2,5} = -\frac{97}{3360} \, , \notag \\
	B_{3,1} &= -\frac{29}{3150} \, , B_{3,2} = -\frac{67}{6300} \, , B_{3,3} = -\frac{17}{1512} \, , B_{3,4} = -\frac{95}{8316} \, , B_{3,5} = -\frac{1351}{118800} \, , \notag \\
	B_{4,1} &= -\frac{53}{25200} \, , B_{4,2} = -\frac{587}{199584} \, , B_{4,3} = -\frac{593}{166320} \, , B_{4,4} = -\frac{4963}{1235520} \, , B_{4,5} = -\frac{5023}{1158300} \, , \notag \\
	B_{5,1} &= -\frac{851}{1746360} \, , B_{5,2} = -\frac{2021}{2522520} \, , B_{5,3} = -\frac{509}{463320} \, , B_{5,4} = -\frac{79}{57915} \, , \notag \\
	B_{6,1} &= -\frac{1381}{12108096} \, , B_{6,2} = -\frac{41}{190080} \, , B_{6,3} = -\frac{134}{405405} \, , \notag \\
	B_{7,1} &= -\frac{349}{12972960} \, , B_{7,2} = -\frac{4759}{82702620} \, , \notag \\
	B_{8,1} &= -\frac{3017}{472586400} \, .
\end{align}
\underline{Coefficients of group $(d)$: $P = g_q^2 + g_{q'}^2$}
\begin{align}
	B_{0,m} &= 0 \notag \\
	B_{1,1} &= -\frac{55}{243} +\frac{16}{81} \, \mathrm{ln} \, x -\frac{1}{27} \, \mathrm{ln}^2 \, x \, , B_{1,2} = -\frac{31}{243} +\frac{13}{81} \, \mathrm{ln} \, x -\frac{1}{27} \, \mathrm{ln}^2 \, x \, , \notag \\
B_{1,3} &= -\frac{2}{27} +\frac{11}{81} \, \mathrm{ln} \, x -\frac{1}{27} \, \mathrm{ln}^2 \, x \, , B_{1,4} = -\frac{13}{324} +\frac{19}{162} \, \mathrm{ln} \, x -\frac{1}{27} \, \mathrm{ln}^2 \, x \, , \notag \\
B_{1,5} &= -\frac{1}{60} +\frac{83}{810} \, \mathrm{ln} \, x -\frac{1}{27} \, \mathrm{ln}^2 \, x \, , \notag \\
	B_{2,1} &= -\frac{91}{7776} +\frac{5}{324} \, \mathrm{ln} \, x -\frac{1}{216} \, \mathrm{ln}^2 \, x \, , B_{2,2} = -\frac{17}{1728} +\frac{1}{54} \, \mathrm{ln} \, x -\frac{1}{144} \, \mathrm{ln}^2 \, x \, , \notag \\
B_{2,3} &= -\frac{1}{144} +\frac{13}{648} \, \mathrm{ln} \, x -\frac{1}{108} \, \mathrm{ln}^2 \, x \, , B_{2,4} = -\frac{19}{5184} +\frac{53}{2592} \, \mathrm{ln} \, x -\frac{5}{432} \, \mathrm{ln}^2 \, x \, , \notag \\
B_{2,5} &= -\frac{1}{3240} +\frac{43}{2160} \, \mathrm{ln} \, x -\frac{1}{72} \, \mathrm{ln}^2 \, x \, , \notag \\
	B_{3,1} &= -\frac{47}{30000} +\frac{37}{13500} \, \mathrm{ln} \, x -\frac{1}{900} \, \mathrm{ln}^2 \, x \, , B_{3,2} = -\frac{119}{67500} +\frac{59}{13500} \, \mathrm{ln} \, x -\frac{1}{450} \, \mathrm{ln}^2 \, x \, , \notag \\
B_{3,3} &= -\frac{1}{675} +\frac{47}{8100} \, \mathrm{ln} \, x -\frac{1}{270} \, \mathrm{ln}^2 \, x \, , B_{3,4} = -\frac{1}{1296} +\frac{37}{5400} \, \mathrm{ln} \, x -\frac{1}{180} \, \mathrm{ln}^2 \, x \, , \notag \\
\displaybreak
B_{3,5} &= \frac{47}{162000} +\frac{199}{27000} \, \mathrm{ln} \, x -\frac{7}{900} \, \mathrm{ln}^2 \, x \, , \label{coeffmixedd} \\
	B_{4,1} &= -\frac{407}{1215000} +\frac{59}{81000} \, \mathrm{ln} \, x -\frac{1}{2700} \, \mathrm{ln}^2 \, x \, , B_{4,2} = -\frac{23}{48600} +\frac{47}{32400} \, \mathrm{ln} \, x -\frac{1}{1080} \, \mathrm{ln}^2 \, x \, , \notag \\
B_{4,3} &= -\frac{1}{2160} +\frac{37}{16200} \, \mathrm{ln} \, x -\frac{1}{540} \, \mathrm{ln}^2 \, x \, , B_{4,4} = -\frac{31}{129600} +\frac{199}{64800} \, \mathrm{ln} \, x -\frac{7}{2160} \, \mathrm{ln}^2 \, x \, , \notag \\
B_{4,5} &= \frac{1}{4320} +\frac{293}{81000} \, \mathrm{ln} \, x -\frac{7}{1350} \, \mathrm{ln}^2 \, x \, , \notag \\
	B_{5,1} &= -\frac{1391}{14586075} +\frac{172}{694575} \, \mathrm{ln} \, x -\frac{1}{6615} \, \mathrm{ln}^2 \, x \, , \notag \\
B_{5,2} &= -\frac{263}{1620675} +\frac{137}{231525} \, \mathrm{ln} \, x -\frac{1}{2205} \, \mathrm{ln}^2 \, x \, , B_{5,3} = -\frac{2}{11025} +\frac{107}{99225} \, \mathrm{ln} \, x -\frac{1}{945} \, \mathrm{ln}^2 \, x \, , \notag \\
B_{5,4} &= -\frac{37}{396900} +\frac{323}{198450} \, \mathrm{ln} \, x -\frac{2}{945} \, \mathrm{ln}^2 \, x \, , \notag \\
	B_{6,1} &= -\frac{365}{11063808} +\frac{59}{592704} \, \mathrm{ln} \, x -\frac{1}{14112} \, \mathrm{ln}^2 \, x \, , \notag \\
B_{6,2} &= -\frac{89}{1354752} +\frac{47}{169344} \, \mathrm{ln} \, x -\frac{1}{4032} \, \mathrm{ln}^2 \, x \, , B_{6,3} = -\frac{1}{12096} +\frac{73}{127008} \, \mathrm{ln} \, x -\frac{1}{1512} \, \mathrm{ln}^2 \, x \, , \notag \\
	B_{7,1} &= -\frac{3247}{246903552} +\frac{155}{3429216} \, \mathrm{ln} \, x -\frac{1}{27216} \, \mathrm{ln}^2 \, x \, , \notag \\
B_{7,2} &= -\frac{463}{15431472} +\frac{247}{1714608} \, \mathrm{ln} \, x -\frac{1}{6804} \, \mathrm{ln}^2 \, x \, , \notag \\
	B_{8,1} &= -\frac{2299}{393660000} +\frac{197}{8748000} \, \mathrm{ln} \, x -\frac{1}{48600} \, \mathrm{ln}^2 \, x \, . \tag{\ref{coeffmixedd}}
\end{align}
\underline{Coefficients of group $(e)$: $P = g_q \, g_{q'}$}
\begin{align}
	B_{0,m} &= 0 \, , \notag \\
	B_{1,1} &= \frac{53}{27} -\frac{4}{3} \, \zeta(3) -\frac{4}{3} \, \mathrm{ln} \, x -\frac{1}{3} \, \mathrm{ln}^2 \, x \, , B_{1,2} = \frac{35}{27} -\frac{4}{3} \, \zeta(3) -\frac{5}{3} \, \mathrm{ln} \, x -\frac{1}{3} \, \mathrm{ln}^2 \, x \, , \notag \\
B_{1,3} &= \frac{20}{27} -\frac{4}{3} \, \zeta(3) -\frac{17}{9} \, \mathrm{ln} \, x -\frac{1}{3} \, \mathrm{ln}^2 \, x \, , B_{1,4} = \frac{29}{108} -\frac{4}{3} \, \zeta(3) -\frac{37}{18} \, \mathrm{ln} \, x -\frac{1}{3} \, \mathrm{ln}^2 \, x \, , \notag \\
B_{1,5} &= -\frac{77}{540} -\frac{4}{3} \, \zeta(3) -\frac{197}{90} \, \mathrm{ln} \, x -\frac{1}{3} \, \mathrm{ln}^2 \, x \, , \notag \\
	B_{2,1} &= -\frac{79}{1296} +\frac{2}{3} \, \zeta(3) +\frac{23}{18} \, \mathrm{ln} \, x -\frac{7}{18} \, \mathrm{ln}^2 \, x \, , B_{2,2} = \frac{473}{864} +\zeta(3) +\frac{55}{36} \, \mathrm{ln} \, x -\frac{7}{12} \, \mathrm{ln}^2 \, x \, , \notag \\
B_{2,3} &= \frac{803}{648} +\frac{4}{3} \, \zeta(3) +\frac{89}{54} \, \mathrm{ln} \, x -\frac{7}{9} \, \mathrm{ln}^2 \, x \, , B_{2,4} = \frac{5083}{2592} +\frac{5}{3} \, \zeta(3) +\frac{361}{216} \, \mathrm{ln} \, x -\frac{35}{36} \, \mathrm{ln}^2 \, x \, , \notag \\
B_{2,5} &= \frac{43}{16} +2 \, \zeta(3) +\frac{97}{60} \, \mathrm{ln} \, x -\frac{7}{6} \, \mathrm{ln}^2 \, x \, , \notag \\
	B_{3,1} &= -\frac{791}{1200} -\frac{2}{5} \, \zeta(3) -\frac{13}{30} \, \mathrm{ln} \, x \, , B_{3,2} = -\frac{307}{200} -\frac{4}{5} \, \zeta(3) -\frac{13}{15} \, \mathrm{ln} \, x \, , \notag \\
B_{3,3} &= -\frac{205}{72} -\frac{4}{3} \, \zeta(3) -\frac{13}{9} \, \mathrm{ln} \, x \, , B_{3,4} = -\frac{667}{144} -2 \, \zeta(3) -\frac{13}{6} \, \mathrm{ln} \, x \, , \notag \\
B_{3,5} &= -\frac{4981}{720} -\frac{14}{5} \, \zeta(3) -\frac{91}{30} \, \mathrm{ln} \, x \, , \notag \\
	B_{4,1} &= \frac{171079}{243000} +\frac{4}{15} \, \zeta(3) +\frac{233}{810} \, \mathrm{ln} \, x -\frac{11}{45} \, \mathrm{ln}^2 \, x \, , \notag \\
B_{4,2} &= \frac{185059}{97200} +\frac{2}{3} \, \zeta(3) +\frac{769}{1620} \, \mathrm{ln} \, x -\frac{11}{18} \, \mathrm{ln}^2 \, x \, , \notag \\
\displaybreak
B_{4,3} &= \frac{192749}{48600} +\frac{4}{3} \, \zeta(3) +\frac{439}{810} \, \mathrm{ln} \, x -\frac{11}{9} \, \mathrm{ln}^2 \, x \, , \label{coeffmixede} \\
B_{4,4} &= \frac{1375583}{194400} +\frac{7}{3} \, \zeta(3) +\frac{1093}{3240} \, \mathrm{ln} \, x -\frac{77}{36} \, \mathrm{ln}^2 \, x \, , \notag \\
B_{4,5} &= \frac{2767561}{243000} +\frac{56}{15} \, \zeta(3) -\frac{1279}{4050} \, \mathrm{ln} \, x -\frac{154}{45} \, \mathrm{ln}^2 \, x \, , \notag \\
	B_{5,1} &= -\frac{16099}{23814} -\frac{4}{21} \, \zeta(3) -\frac{37}{567} \, \mathrm{ln} \, x \, , B_{5,2} = -\frac{8179}{3969} -\frac{4}{7} \, \zeta(3) -\frac{37}{189} \, \mathrm{ln} \, x \, , \notag \\
B_{5,3} &= -\frac{8290}{1701} -\frac{4}{3} \, \zeta(3) -\frac{37}{81} \, \mathrm{ln} \, x \, , B_{5,4} = -\frac{67097}{6804} -\frac{8}{3} \, \zeta(3) -\frac{74}{81} \, \mathrm{ln} \, x \, , \notag \\
	B_{6,1} &= \frac{529489097}{889056000} +\frac{1}{7} \, \zeta(3) +\frac{689}{15120} \, \mathrm{ln} \, x -\frac{143}{840} \, \mathrm{ln}^2 \, x \, , \notag \\
B_{6,2} &= \frac{535276697}{254016000} +\frac{1}{2} \, \zeta(3) -\frac{65}{6048} \, \mathrm{ln} \, x -\frac{143}{240} \, \mathrm{ln}^2 \, x \, , \notag \\
B_{6,3} &= \frac{534935447}{95256000} +\frac{4}{3} \, \zeta(3) -\frac{9659}{22680} \, \mathrm{ln} \, x -\frac{143}{90} \, \mathrm{ln}^2 \, x \, , \notag \\
	B_{7,1} &= -\frac{259882013}{489888000} -\frac{1}{9} \, \zeta(3) +\frac{763}{19440} \, \mathrm{ln} \, x -\frac{7}{1080} \, \mathrm{ln}^2 \, x \, , \notag \\
B_{7,2} &= -\frac{257478563}{122472000} -\frac{4}{9} \, \zeta(3) +\frac{1463}{9720} \, \mathrm{ln} \, x -\frac{7}{270} \, \mathrm{ln}^2 \, x \, , \notag \\
	B_{8,1} &= \frac{13910828791}{30005640000} +\frac{4}{45} \, \zeta(3) -\frac{52699}{1701000} \, \mathrm{ln} \, x -\frac{403}{3150} \, \mathrm{ln}^2 \, x \, . \tag{\ref{coeffmixede}}
\end{align}

\end{appendix}

\phantomsection
\addcontentsline{toc}{section}{References}
\bibliographystyle{utphys}
\bibliography{arxiv}

\end{document}